\documentclass[twocolumn,twocolappendix]{aastex701}
\usepackage{graphicx}
\usepackage{amsmath}
\usepackage{array}
\usepackage{enumitem}
\usepackage{soul}
\usepackage{tikz}
\usetikzlibrary{arrows.meta}

\graphicspath{{plots/}}

\newcommand{\Msun}{{\rm M}_{\odot}}
\newcommand{\Lenstool}{\text{\sc{Lenstool}}}

\begin{document}


\newcommand{\SimulatedLineOfSightClassChangeClusterBalancedMeanFraction}{0.0244}
\newcommand{\SimulatedLineOfSightClassChangeClusterBalancedMeanPercent}{2.4}
\newcommand{\ObservedSampleTotalMemberCount}{1203}
\newcommand{\RepresentativeAnalogSampleTotalMemberCount}{1203}
\newcommand{\ObservedClusterCount}{6}
\newcommand{\ObservedClustersAboveSimulatedMeanPerturberFractionCount}{5}
\newcommand{\ObservedClustersAboveSimulatedMedianPerturberFractionCount}{5}
\newcommand{\SimulatedAnalogsPerObservedClusterMinCount}{10}
\newcommand{\SimulatedAnalogsPerObservedClusterMaxCount}{10}
\newcommand{\SimulatedAnalogsPerObservedClusterCount}{10}
\newcommand{\SimulatedWithoutLineOfSightAnalogCount}{60}
\newcommand{\RepresentativeAnalogCount}{6}
\newcommand{\ObservedVsSimulatedWithoutLineOfSightGlobalCompositionExactMatchedPermutationGaussianEquivalentZ}{+3.48}
\newcommand{\ObservedVsSimulatedWithoutLineOfSightGlobalCompositionExactMatchedPermutationPValue}{5.0 \times 10^{-4}}

\newcommand{\ObservedSampleIntrinsicLensCount}{137}
\newcommand{\ObservedSampleIntrinsicLensPooledFraction}{0.114}
\newcommand{\ObservedSampleIntrinsicLensPooledPercent}{11.4}
\newcommand{\RepresentativeAnalogSampleIntrinsicLensCount}{87}
\newcommand{\RepresentativeAnalogSampleIntrinsicLensPooledFraction}{0.072}
\newcommand{\RepresentativeAnalogSampleIntrinsicLensPooledPercent}{7.2}
\newcommand{\ObservedIntrinsicLensClusterBalancedMeanFraction}{0.129}
\newcommand{\ObservedIntrinsicLensClusterBalancedMeanPercent}{12.9}
\newcommand{\SimulatedWithoutLineOfSightIntrinsicLensClusterBalancedMeanFraction}{0.090}
\newcommand{\SimulatedWithoutLineOfSightIntrinsicLensClusterBalancedMeanPercent}{9.0}
\newcommand{\SimulatedWithLineOfSightIntrinsicLensClusterBalancedMeanFraction}{0.087}
\newcommand{\SimulatedWithLineOfSightIntrinsicLensClusterBalancedMeanPercent}{8.7}
\newcommand{\ObservedToSimulatedWithoutLineOfSightIntrinsicLensClusterBalancedFractionRatio}{1.44}
\newcommand{\ObservedMinusSimulatedWithoutLineOfSightIntrinsicLensClusterBalancedPercentagePointDifference}{+3.9}
\newcommand{\ObservedVsSimulatedWithoutLineOfSightIntrinsicLensExactMatchedPermutationGaussianEquivalentZ}{+2.19}
\newcommand{\ObservedVsSimulatedWithoutLineOfSightIntrinsicLensExactMatchedPermutationRawPValue}{0.03}
\newcommand{\ObservedVsSimulatedWithoutLineOfSightIntrinsicLensExactMatchedPermutationHolmAdjustedPValue}{0.04}
\newcommand{\ObservedVsSimulatedWithoutLineOfSightIntrinsicLensExactStratifiedRankGaussianEquivalentZ}{+1.47}
\newcommand{\ObservedVsSimulatedWithoutLineOfSightIntrinsicLensExactStratifiedRankOneSidedPValue}{0.07}
\newcommand{\ObservedMinusSimulatedWithoutLineOfSightIntrinsicLensAnalogScatterStandardDeviationEffectSize}{+1.13}

\newcommand{\ObservedSampleBoostedLensCount}{197}
\newcommand{\ObservedSampleBoostedLensPooledFraction}{0.164}
\newcommand{\ObservedSampleBoostedLensPooledPercent}{16.4}
\newcommand{\RepresentativeAnalogSampleBoostedLensCount}{131}
\newcommand{\RepresentativeAnalogSampleBoostedLensPooledFraction}{0.109}
\newcommand{\RepresentativeAnalogSampleBoostedLensPooledPercent}{10.9}
\newcommand{\ObservedBoostedLensClusterBalancedMeanFraction}{0.169}
\newcommand{\ObservedBoostedLensClusterBalancedMeanPercent}{16.9}
\newcommand{\SimulatedWithoutLineOfSightBoostedLensClusterBalancedMeanFraction}{0.104}
\newcommand{\SimulatedWithoutLineOfSightBoostedLensClusterBalancedMeanPercent}{10.4}
\newcommand{\SimulatedWithLineOfSightBoostedLensClusterBalancedMeanFraction}{0.111}
\newcommand{\SimulatedWithLineOfSightBoostedLensClusterBalancedMeanPercent}{11.1}
\newcommand{\ObservedToSimulatedWithoutLineOfSightBoostedLensClusterBalancedFractionRatio}{1.62}
\newcommand{\ObservedMinusSimulatedWithoutLineOfSightBoostedLensClusterBalancedPercentagePointDifference}{+6.5}
\newcommand{\ObservedVsSimulatedWithoutLineOfSightBoostedLensExactMatchedPermutationGaussianEquivalentZ}{+3.57}
\newcommand{\ObservedVsSimulatedWithoutLineOfSightBoostedLensExactMatchedPermutationRawPValue}{3.6 \times 10^{-4}}
\newcommand{\ObservedVsSimulatedWithoutLineOfSightBoostedLensExactMatchedPermutationHolmAdjustedPValue}{1.4 \times 10^{-3}}
\newcommand{\ObservedVsSimulatedWithoutLineOfSightBoostedLensExactStratifiedRankGaussianEquivalentZ}{+2.77}
\newcommand{\ObservedVsSimulatedWithoutLineOfSightBoostedLensExactStratifiedRankOneSidedPValue}{2.8 \times 10^{-3}}
\newcommand{\ObservedMinusSimulatedWithoutLineOfSightBoostedLensAnalogScatterStandardDeviationEffectSize}{+1.82}

\newcommand{\ObservedSampleMajorPerturberCount}{76}
\newcommand{\ObservedSampleMajorPerturberPooledFraction}{0.063}
\newcommand{\ObservedSampleMajorPerturberPooledPercent}{6.3}
\newcommand{\RepresentativeAnalogSampleMajorPerturberCount}{47}
\newcommand{\RepresentativeAnalogSampleMajorPerturberPooledFraction}{0.039}
\newcommand{\RepresentativeAnalogSampleMajorPerturberPooledPercent}{3.9}
\newcommand{\ObservedMajorPerturberClusterBalancedMeanFraction}{0.062}
\newcommand{\ObservedMajorPerturberClusterBalancedMeanPercent}{6.2}
\newcommand{\SimulatedWithoutLineOfSightMajorPerturberClusterBalancedMeanFraction}{0.037}
\newcommand{\SimulatedWithoutLineOfSightMajorPerturberClusterBalancedMeanPercent}{3.7}
\newcommand{\SimulatedWithLineOfSightMajorPerturberClusterBalancedMeanFraction}{0.040}
\newcommand{\SimulatedWithLineOfSightMajorPerturberClusterBalancedMeanPercent}{4.0}
\newcommand{\ObservedToSimulatedWithoutLineOfSightMajorPerturberClusterBalancedFractionRatio}{1.66}
\newcommand{\ObservedMinusSimulatedWithoutLineOfSightMajorPerturberClusterBalancedPercentagePointDifference}{+2.4}
\newcommand{\ObservedVsSimulatedWithoutLineOfSightMajorPerturberExactMatchedPermutationGaussianEquivalentZ}{+2.30}
\newcommand{\ObservedVsSimulatedWithoutLineOfSightMajorPerturberExactMatchedPermutationRawPValue}{0.02}
\newcommand{\ObservedVsSimulatedWithoutLineOfSightMajorPerturberExactMatchedPermutationHolmAdjustedPValue}{0.04}
\newcommand{\ObservedVsSimulatedWithoutLineOfSightMajorPerturberExactStratifiedRankGaussianEquivalentZ}{+1.74}
\newcommand{\ObservedVsSimulatedWithoutLineOfSightMajorPerturberExactStratifiedRankOneSidedPValue}{0.04}
\newcommand{\ObservedMinusSimulatedWithoutLineOfSightMajorPerturberAnalogScatterStandardDeviationEffectSize}{+1.19}

\newcommand{\ObservedSampleMinorPerturberCount}{358}
\newcommand{\ObservedSampleMinorPerturberPooledFraction}{0.298}
\newcommand{\ObservedSampleMinorPerturberPooledPercent}{29.8}
\newcommand{\RepresentativeAnalogSampleMinorPerturberCount}{223}
\newcommand{\RepresentativeAnalogSampleMinorPerturberPooledFraction}{0.185}
\newcommand{\RepresentativeAnalogSampleMinorPerturberPooledPercent}{18.5}
\newcommand{\ObservedMinorPerturberClusterBalancedMeanFraction}{0.275}
\newcommand{\ObservedMinorPerturberClusterBalancedMeanPercent}{27.5}
\newcommand{\SimulatedWithoutLineOfSightMinorPerturberClusterBalancedMeanFraction}{0.178}
\newcommand{\SimulatedWithoutLineOfSightMinorPerturberClusterBalancedMeanPercent}{17.8}
\newcommand{\SimulatedWithLineOfSightMinorPerturberClusterBalancedMeanFraction}{0.186}
\newcommand{\SimulatedWithLineOfSightMinorPerturberClusterBalancedMeanPercent}{18.6}
\newcommand{\ObservedToSimulatedWithoutLineOfSightMinorPerturberClusterBalancedFractionRatio}{1.54}
\newcommand{\ObservedMinusSimulatedWithoutLineOfSightMinorPerturberClusterBalancedPercentagePointDifference}{+9.7}
\newcommand{\ObservedVsSimulatedWithoutLineOfSightMinorPerturberExactMatchedPermutationGaussianEquivalentZ}{+2.59}
\newcommand{\ObservedVsSimulatedWithoutLineOfSightMinorPerturberExactMatchedPermutationRawPValue}{9.6 \times 10^{-3}}
\newcommand{\ObservedVsSimulatedWithoutLineOfSightMinorPerturberExactMatchedPermutationHolmAdjustedPValue}{0.03}
\newcommand{\ObservedVsSimulatedWithoutLineOfSightMinorPerturberExactStratifiedRankGaussianEquivalentZ}{+1.74}
\newcommand{\ObservedVsSimulatedWithoutLineOfSightMinorPerturberExactStratifiedRankOneSidedPValue}{0.04}
\newcommand{\ObservedMinusSimulatedWithoutLineOfSightMinorPerturberAnalogScatterStandardDeviationEffectSize}{+1.16}

\newcommand{\ObservedSampleInertMemberCount}{435}
\newcommand{\ObservedSampleInertMemberPooledFraction}{0.362}
\newcommand{\ObservedSampleInertMemberPooledPercent}{36.2}
\newcommand{\RepresentativeAnalogSampleInertMemberCount}{715}
\newcommand{\RepresentativeAnalogSampleInertMemberPooledFraction}{0.594}
\newcommand{\RepresentativeAnalogSampleInertMemberPooledPercent}{59.4}
\newcommand{\ObservedInertMemberClusterBalancedMeanFraction}{0.364}
\newcommand{\ObservedInertMemberClusterBalancedMeanPercent}{36.4}
\newcommand{\SimulatedWithoutLineOfSightInertMemberClusterBalancedMeanFraction}{0.590}
\newcommand{\SimulatedWithoutLineOfSightInertMemberClusterBalancedMeanPercent}{59.0}
\newcommand{\SimulatedWithLineOfSightInertMemberClusterBalancedMeanFraction}{0.576}
\newcommand{\SimulatedWithLineOfSightInertMemberClusterBalancedMeanPercent}{57.6}
\newcommand{\ObservedToSimulatedWithoutLineOfSightInertMemberClusterBalancedFractionRatio}{0.62}
\newcommand{\ObservedMinusSimulatedWithoutLineOfSightInertMemberClusterBalancedPercentagePointDifference}{-22.6}
\newcommand{\ObservedVsSimulatedWithoutLineOfSightInertMemberExactMatchedPermutationGaussianEquivalentZ}{-3.97}
\newcommand{\ObservedVsSimulatedWithoutLineOfSightInertMemberExactMatchedPermutationRawPValue}{7.2 \times 10^{-5}}
\newcommand{\ObservedVsSimulatedWithoutLineOfSightInertMemberExactMatchedPermutationHolmAdjustedPValue}{3.6 \times 10^{-4}}
\newcommand{\ObservedVsSimulatedWithoutLineOfSightInertMemberExactStratifiedRankGaussianEquivalentZ}{-3.68}
\newcommand{\ObservedVsSimulatedWithoutLineOfSightInertMemberExactStratifiedRankOneSidedPValue}{1.2 \times 10^{-4}}
\newcommand{\ObservedMinusSimulatedWithoutLineOfSightInertMemberAnalogScatterStandardDeviationEffectSize}{-2.09}

\newcommand{\ObservedSampleLensesCount}{334}
\newcommand{\ObservedSampleLensesPooledFraction}{0.278}
\newcommand{\ObservedSampleLensesPooledPercent}{27.8}
\newcommand{\RepresentativeAnalogSampleLensesCount}{218}
\newcommand{\RepresentativeAnalogSampleLensesPooledFraction}{0.181}
\newcommand{\RepresentativeAnalogSampleLensesPooledPercent}{18.1}
\newcommand{\ObservedLensesClusterBalancedMeanFraction}{0.299}
\newcommand{\ObservedLensesClusterBalancedMeanPercent}{29.9}
\newcommand{\SimulatedWithoutLineOfSightLensesClusterBalancedMeanFraction}{0.194}
\newcommand{\SimulatedWithoutLineOfSightLensesClusterBalancedMeanPercent}{19.4}
\newcommand{\SimulatedWithLineOfSightLensesClusterBalancedMeanFraction}{0.199}
\newcommand{\SimulatedWithLineOfSightLensesClusterBalancedMeanPercent}{19.9}
\newcommand{\ObservedToSimulatedWithoutLineOfSightLensesClusterBalancedFractionRatio}{1.54}
\newcommand{\ObservedMinusSimulatedWithoutLineOfSightLensesClusterBalancedPercentagePointDifference}{+10.4}

\newcommand{\ObservedSamplePerturbersCount}{434}
\newcommand{\ObservedSamplePerturbersPooledFraction}{0.361}
\newcommand{\ObservedSamplePerturbersPooledPercent}{36.1}
\newcommand{\RepresentativeAnalogSamplePerturbersCount}{270}
\newcommand{\RepresentativeAnalogSamplePerturbersPooledFraction}{0.224}
\newcommand{\RepresentativeAnalogSamplePerturbersPooledPercent}{22.4}
\newcommand{\ObservedPerturbersClusterBalancedMeanFraction}{0.337}
\newcommand{\ObservedPerturbersClusterBalancedMeanPercent}{33.7}
\newcommand{\SimulatedWithoutLineOfSightPerturbersClusterBalancedMeanFraction}{0.216}
\newcommand{\SimulatedWithoutLineOfSightPerturbersClusterBalancedMeanPercent}{21.6}
\newcommand{\SimulatedWithLineOfSightPerturbersClusterBalancedMeanFraction}{0.226}
\newcommand{\SimulatedWithLineOfSightPerturbersClusterBalancedMeanPercent}{22.6}
\newcommand{\ObservedToSimulatedWithoutLineOfSightPerturbersClusterBalancedFractionRatio}{1.56}
\newcommand{\ObservedMinusSimulatedWithoutLineOfSightPerturbersClusterBalancedPercentagePointDifference}{+12.1}

\newcommand{\ObservedClusterCriticalRegionClusterBalancedMeanFraction}{0.337}
\newcommand{\ObservedClusterCriticalRegionClusterBalancedMeanPercent}{33.7}
\newcommand{\SimulatedWithoutLineOfSightClusterCriticalRegionClusterBalancedMeanFraction}{0.216}
\newcommand{\SimulatedWithoutLineOfSightClusterCriticalRegionClusterBalancedMeanPercent}{21.6}
\newcommand{\SimulatedWithLineOfSightClusterCriticalRegionClusterBalancedMeanFraction}{0.226}
\newcommand{\SimulatedWithLineOfSightClusterCriticalRegionClusterBalancedMeanPercent}{22.6}
\newcommand{\SimulatedToObservedWithoutLineOfSightClusterCriticalRegionClusterBalancedFractionRatio}{0.64}

\newcommand{\ObservedClusterNoncriticalRegionClusterBalancedMeanFraction}{0.663}
\newcommand{\ObservedClusterNoncriticalRegionClusterBalancedMeanPercent}{66.3}
\newcommand{\SimulatedWithoutLineOfSightClusterNoncriticalRegionClusterBalancedMeanFraction}{0.784}
\newcommand{\SimulatedWithoutLineOfSightClusterNoncriticalRegionClusterBalancedMeanPercent}{78.4}
\newcommand{\SimulatedWithLineOfSightClusterNoncriticalRegionClusterBalancedMeanFraction}{0.774}
\newcommand{\SimulatedWithLineOfSightClusterNoncriticalRegionClusterBalancedMeanPercent}{77.4}
\newcommand{\SimulatedToObservedWithoutLineOfSightClusterNoncriticalRegionClusterBalancedFractionRatio}{1.18}

\newcommand{\ObservedMajorPerturberWithinClusterCriticalRegionClusterBalancedMeanFraction}{0.214}
\newcommand{\ObservedMajorPerturberWithinClusterCriticalRegionClusterBalancedMeanPercent}{21.4}
\newcommand{\SimulatedWithoutLineOfSightMajorPerturberWithinClusterCriticalRegionClusterBalancedMeanFraction}{0.175}
\newcommand{\SimulatedWithoutLineOfSightMajorPerturberWithinClusterCriticalRegionClusterBalancedMeanPercent}{17.5}
\newcommand{\SimulatedWithLineOfSightMajorPerturberWithinClusterCriticalRegionClusterBalancedMeanFraction}{0.179}
\newcommand{\SimulatedWithLineOfSightMajorPerturberWithinClusterCriticalRegionClusterBalancedMeanPercent}{17.9}
\newcommand{\SimulatedToObservedWithoutLineOfSightMajorPerturberWithinClusterCriticalRegionClusterBalancedFractionRatio}{0.82}

\newcommand{\ObservedMinorPerturberWithinClusterCriticalRegionClusterBalancedMeanFraction}{0.786}
\newcommand{\ObservedMinorPerturberWithinClusterCriticalRegionClusterBalancedMeanPercent}{78.6}
\newcommand{\SimulatedWithoutLineOfSightMinorPerturberWithinClusterCriticalRegionClusterBalancedMeanFraction}{0.825}
\newcommand{\SimulatedWithoutLineOfSightMinorPerturberWithinClusterCriticalRegionClusterBalancedMeanPercent}{82.5}
\newcommand{\SimulatedWithLineOfSightMinorPerturberWithinClusterCriticalRegionClusterBalancedMeanFraction}{0.821}
\newcommand{\SimulatedWithLineOfSightMinorPerturberWithinClusterCriticalRegionClusterBalancedMeanPercent}{82.1}
\newcommand{\SimulatedToObservedWithoutLineOfSightMinorPerturberWithinClusterCriticalRegionClusterBalancedFractionRatio}{1.05}

\newcommand{\ObservedIntrinsicLensWithinClusterNoncriticalRegionClusterBalancedMeanFraction}{0.186}
\newcommand{\ObservedIntrinsicLensWithinClusterNoncriticalRegionClusterBalancedMeanPercent}{18.6}
\newcommand{\SimulatedWithoutLineOfSightIntrinsicLensWithinClusterNoncriticalRegionClusterBalancedMeanFraction}{0.115}
\newcommand{\SimulatedWithoutLineOfSightIntrinsicLensWithinClusterNoncriticalRegionClusterBalancedMeanPercent}{11.5}
\newcommand{\SimulatedWithLineOfSightIntrinsicLensWithinClusterNoncriticalRegionClusterBalancedMeanFraction}{0.113}
\newcommand{\SimulatedWithLineOfSightIntrinsicLensWithinClusterNoncriticalRegionClusterBalancedMeanPercent}{11.3}
\newcommand{\SimulatedToObservedWithoutLineOfSightIntrinsicLensWithinClusterNoncriticalRegionClusterBalancedFractionRatio}{0.62}

\newcommand{\ObservedBoostedLensWithinClusterNoncriticalRegionClusterBalancedMeanFraction}{0.258}
\newcommand{\ObservedBoostedLensWithinClusterNoncriticalRegionClusterBalancedMeanPercent}{25.8}
\newcommand{\SimulatedWithoutLineOfSightBoostedLensWithinClusterNoncriticalRegionClusterBalancedMeanFraction}{0.135}
\newcommand{\SimulatedWithoutLineOfSightBoostedLensWithinClusterNoncriticalRegionClusterBalancedMeanPercent}{13.5}
\newcommand{\SimulatedWithLineOfSightBoostedLensWithinClusterNoncriticalRegionClusterBalancedMeanFraction}{0.147}
\newcommand{\SimulatedWithLineOfSightBoostedLensWithinClusterNoncriticalRegionClusterBalancedMeanPercent}{14.7}
\newcommand{\SimulatedToObservedWithoutLineOfSightBoostedLensWithinClusterNoncriticalRegionClusterBalancedFractionRatio}{0.52}

\newcommand{\ObservedInertMemberWithinClusterNoncriticalRegionClusterBalancedMeanFraction}{0.556}
\newcommand{\ObservedInertMemberWithinClusterNoncriticalRegionClusterBalancedMeanPercent}{55.6}
\newcommand{\SimulatedWithoutLineOfSightInertMemberWithinClusterNoncriticalRegionClusterBalancedMeanFraction}{0.750}
\newcommand{\SimulatedWithoutLineOfSightInertMemberWithinClusterNoncriticalRegionClusterBalancedMeanPercent}{75.0}
\newcommand{\SimulatedWithLineOfSightInertMemberWithinClusterNoncriticalRegionClusterBalancedMeanFraction}{0.740}
\newcommand{\SimulatedWithLineOfSightInertMemberWithinClusterNoncriticalRegionClusterBalancedMeanPercent}{74.0}
\newcommand{\SimulatedToObservedWithoutLineOfSightInertMemberWithinClusterNoncriticalRegionClusterBalancedFractionRatio}{1.35}

\newcommand{\HessianRadiusComparisonMedianDifferenceMin}{2.1}
\newcommand{\HessianRadiusComparisonMedianDifferenceMax}{3.2}
\newcommand{\HessianRadiusComparisonClassificationAgreementMin}{98.5}
\newcommand{\HessianRadiusComparisonClassificationAgreementMax}{99.5}


\newcommand{\CoreCollapsedSIDMPlummerScale}{0.089^{+0.148}_{-0.048}\,\mathrm{kpc}}
\newcommand{\CoreCollapsedSIDMResponseTransitionMass}{(2.50^{+0.46}_{-0.62})\times 10^{10}}
\newcommand{\CoreCollapsedSIDMResponseLogMassWidth}{0.396^{+0.050}_{-0.035}}

\newcommand{\SimulatedWithCoreCollapsedSIDMClusterCriticalRegionClusterBalancedMeanFraction}{0.217}
\newcommand{\SimulatedWithCoreCollapsedSIDMClusterCriticalRegionClusterBalancedMeanPercent}{21.7}
\newcommand{\SimulatedWithCoreCollapsedSIDMClusterNoncriticalRegionClusterBalancedMeanFraction}{0.783}
\newcommand{\SimulatedWithCoreCollapsedSIDMClusterNoncriticalRegionClusterBalancedMeanPercent}{78.3}
\newcommand{\SimulatedWithCoreCollapsedSIDMMajorPerturberWithinClusterCriticalRegionClusterBalancedMeanFraction}{0.224}
\newcommand{\SimulatedWithCoreCollapsedSIDMMajorPerturberWithinClusterCriticalRegionClusterBalancedMeanPercent}{22.4}
\newcommand{\SimulatedWithCoreCollapsedSIDMMinorPerturberWithinClusterCriticalRegionClusterBalancedMeanFraction}{0.776}
\newcommand{\SimulatedWithCoreCollapsedSIDMMinorPerturberWithinClusterCriticalRegionClusterBalancedMeanPercent}{77.6}
\newcommand{\SimulatedWithCoreCollapsedSIDMIntrinsicLensWithinClusterNoncriticalRegionClusterBalancedMeanFraction}{0.189}
\newcommand{\SimulatedWithCoreCollapsedSIDMIntrinsicLensWithinClusterNoncriticalRegionClusterBalancedMeanPercent}{18.9}
\newcommand{\SimulatedWithCoreCollapsedSIDMBoostedLensWithinClusterNoncriticalRegionClusterBalancedMeanFraction}{0.244}
\newcommand{\SimulatedWithCoreCollapsedSIDMBoostedLensWithinClusterNoncriticalRegionClusterBalancedMeanPercent}{24.4}
\newcommand{\SimulatedWithCoreCollapsedSIDMInertMemberWithinClusterNoncriticalRegionClusterBalancedMeanFraction}{0.567}
\newcommand{\SimulatedWithCoreCollapsedSIDMInertMemberWithinClusterNoncriticalRegionClusterBalancedMeanPercent}{56.7}
\newcommand{\CoreCollapsedSIDMBootstrapSampleCount}{1,000}
\newcommand{\CoreCollapsedSIDMResponseTransitionMassBootstrapMedian}{2.50 \times 10^{10}}
\newcommand{\CoreCollapsedSIDMResponseTransitionMassBootstrapLower}{1.88 \times 10^{10}}
\newcommand{\CoreCollapsedSIDMResponseTransitionMassBootstrapUpper}{2.97 \times 10^{10}}
\newcommand{\CoreCollapsedSIDMResponseLogMassWidthBootstrapMedian}{0.396}
\newcommand{\CoreCollapsedSIDMResponseLogMassWidthBootstrapLower}{0.361}
\newcommand{\CoreCollapsedSIDMResponseLogMassWidthBootstrapUpper}{0.446}
\newcommand{\CoreCollapsedSIDMPlummerScaleBootstrapMedian}{0.089\,\mathrm{kpc}}
\newcommand{\CoreCollapsedSIDMPlummerScaleBootstrapLower}{0.041\,\mathrm{kpc}}
\newcommand{\CoreCollapsedSIDMPlummerScaleBootstrapUpper}{0.238\,\mathrm{kpc}}
\newcommand{\CoreCollapsedSIDMResponseTransitionMassBestFit}{2.5 \times 10^{10}}
\newcommand{\CoreCollapsedSIDMResponseLogMassWidthBestFit}{0.39}
\newcommand{\CoreCollapsedSIDMPlummerScaleBestFit}{0.124\,\mathrm{kpc}}


\title{Where Cluster Galaxies Reside and How Strongly They Lens}
\shortauthors{Dutra}

\author[0000-0001-7040-4930]{Isaque Dutra}
\affiliation{Department of Physics, Yale University, New Haven, CT 06511, USA}
\email{isaque.dutra@yale.edu}
\correspondingauthor{Isaque Dutra}

\author[0000-0002-5554-8896]{Priyamvada Natarajan}
\affiliation{Department of Astronomy, Yale University, New Haven, CT 06511, USA}
\affiliation{Department of Physics, Yale University, New Haven, CT 06511, USA}
\email{priyamvada.natarajan@yale.edu}

\author[0000-0002-3400-6991]{Cian Roche}
\affiliation{Department of Physics and MIT Kavli Institute for Astrophysics and Space Research,\\ Massachusetts Institute of Technology, Cambridge, MA 02139, USA}
\email{roche@mit.edu}

\begin{abstract}
The efficiency of galaxy-galaxy strong lensing (GGSL) in clusters depends jointly on the inner mass distributions of member galaxies and their locations within the cluster potential. Observed GGSL probabilities exceed $\Lambda$CDM predictions by nearly an order of magnitude, but the relative roles of internal structure and environment remain unclear. We disentangle these effects by comparing HST strong-lensing clusters with TNG-Cluster analogs at similar lens redshifts, matched by their primary Einstein radius. For each member, we measure its lensing strength in isolation and when embedded in the cluster field. Members in cluster-critical regions are classified as Major or Minor Perturbers depending on whether they strongly lens in isolation; those in subcritical regions are classified as Intrinsic Lenses, Boosted Lenses, or Inert Members according to whether they lens independently, only with environmental assistance, or not at all. Observed clusters contain \ObservedClusterCriticalRegionClusterBalancedMeanPercent\% of selected members in cluster-critical regions, compared with \SimulatedWithoutLineOfSightClusterCriticalRegionClusterBalancedMeanPercent\% in TNG-Cluster, while their Major-to-Minor ratios are similar. Outside these regions, observations contain substantially more Intrinsic and Boosted Lenses and fewer Inert Members, with Boosted Lens fractions of \ObservedBoostedLensWithinClusterNoncriticalRegionClusterBalancedMeanPercent\% and \SimulatedWithoutLineOfSightBoostedLensWithinClusterNoncriticalRegionClusterBalancedMeanPercent\%, respectively. The discrepancy persists at fixed subhalo mass and after including line-of-sight structure. A mass-conserving contraction mimicking core-collapsed self-interacting dark matter largely recovers the conditional lensing fractions but not cluster-critical occupancy. The GGSL discrepancy therefore reflects both insufficient central concentration in simulated members and differences in their locations within the cluster lensing field.
\end{abstract}


\section{Introduction} \label{sec:intro}

According to the standard $\Lambda$CDM paradigm, the hierarchical growth of structure culminates in galaxy clusters, the most massive virialized systems in the Universe. Their dense cores provide powerful laboratories for testing the interplay between dark matter and baryons over a wide range of scales \citep[see reviews by][]{Kneib2011, Umetsu2020, Natarajan2024}. Strong gravitational lensing probes both the smooth cluster-scale potential and the population of member galaxy subhalos, making it simultaneously sensitive to the global mass distribution and to the inner mass profiles of individual galaxies \citep{Natarajan2004, Limousin2005, Natarajan2007, Natarajan2009, Richard2014, Jauzac2015, Natarajan2024}.

Cluster strong lensing has long provided tests of $\Lambda$CDM. Early comparisons between observed and simulated giant-arc abundances revealed the so-called ``arc statistics problem'' \citep{Bartelmann1998, Luppino1999, Gladders2003}. Subsequent work showed that arc production depends sensitively on cluster selection, triaxiality, mergers, central galaxies, source redshift distributions, and line-of-sight structure, leaving the magnitude of that discrepancy under debate \citep{Meneghetti2003, Dalal2004, Torri2004, Wambsganss2004, Horesh2011, Meneghetti2011, Meneghetti2013}. 

A distinct tension has since emerged on galaxy scales within clusters. Galaxy-Galaxy Strong Lensing (GGSL) occurs when an individual cluster member produces a secondary critical curve, distinct from the cluster-scale critical curve. The GGSL probability is set by the total source-plane area enclosed by the associated secondary caustics. Because galaxy-scale lenses typically reside where the host cluster provides substantial convergence and shear while remaining locally subcritical, GGSL depends jointly on the inner mass distribution of the member galaxy and the lensing enhancement supplied by its environment.

Observed GGSL probabilities in several massive strong-lensing clusters exceed predictions from state-of-the-art hydrodynamical $\Lambda$CDM simulations by more than one order of magnitude \citep{Meneghetti2020,Meneghetti2022,Meneghetti2023,Betts2026:SLICE_SPTJ1150}. Although baryonic physics can substantially alter the centers of cluster galaxies, the modifications explored so far do not account for this excess. In particular, \citet{Tokayer2024} explored a broad range of baryonic modifications, including extreme adiabatic-contraction models, and showed that they do not recover the observed GGSL signal. \citet{Dutra2025} likewise showed that reproducing the observed GGSL probability requires subhalos with substantially steeper inner density profiles than those produced in standard hydrodynamical CDM simulations. These results point to a missing population of cluster members with sufficiently high projected central densities in simulations. 

GGSL in simulated clusters can therefore be low either because member galaxies are too weak to lens efficiently or because they occupy the wrong parts of the host lensing field \citep{RagagninMeneghettiNatarajan2022}. Existing comparisons based on the total GGSL probability do not cleanly separate these possibilities. This distinction is especially important, as the cluster environment can transform the lensing behavior of an otherwise subcritical member galaxy. A member far from the cluster core may need to be intrinsically critical to produce GGSL, whereas a similar galaxy embedded in a stronger local convergence and shear field may cross the critical threshold only after the host contribution is included \citep{BlandfordKochanek1987,SchneiderEhlersFalco1992,Meneghetti2003,King2007}. The GGSL signal, therefore, depends not only on the members' internal structure but also on where those members reside relative to the cluster critical structure. 

Several observations already suggest that these two ingredients should be considered separately. Lensing measurements find that the resolved subhalo mass function in cluster cores is broadly consistent with $\Lambda$CDM simulations over the mass range currently probed \citep{Natarajan2007,Natarajan+2017}, arguing against subhalo abundance as a source of the tension. At the same time, the projected radial distributions of observed and simulated cluster members differ significantly \citep{Natarajan+2017,Natarajan+2026}. Thus, simulations may contain approximately the correct number of subhalos while placing too few of them in regions where the host cluster can enhance their lensing efficiency or while producing members whose inner density profiles are too shallow for that enhancement to make them critical. 

The importance of the host cluster is also evident from \citet{Roche2026GGSL} who found an approximate scaling $P_{\rm GGSL}\propto\theta_{\rm E,p}^{2}$, where $\theta_{\rm E,p}$ is the Einstein radius of the primary cluster-scale tangential critical curve. Matching observed and simulated clusters in both mass and primary Einstein radius substantially reduces the discrepancy in GGSL probability; however, the remaining difference depends on the field of view. This demonstrates that comparisons of member structure must control for the lensing strength of the host. Matching the primary Einstein radius, however, does not guarantee that individual galaxies experience the same local convergence and shear, nor that their internal mass distributions are equivalent. 

A mechanism that increases the central densities of cluster members while leaving their outer tidal radius structure approximately unchanged could address the internal-structure component of the discrepancy. Gravothermal evolution in self-interacting dark matter (SIDM) provides one such possibility. While dark-matter scattering initially turns a central cusp in the inner density profile into an approximately isothermal core, further evolution can cause systems to enter gravothermal collapse \citep{SpergelSteinhardt2000,Adhikari2025}. As the core approaches thermalization, the temperature gradient reverses, and heat begins to flow outward. Because the self-gravitating core has negative heat capacity, this energy loss causes it to contract and heat further, eventually producing a compact central cusp \citep{Balberg2002,Koda2011}. Whether a cluster member reaches this late collapse phase, however, depends on both its internal structure and environmental processing via its orbit in the host. Interactions that transport heat within the member occur at relative velocities comparable to the member’s internal velocity dispersion, whereas interactions with host particles that drive evaporation occur at the higher relative velocity set by the member’s cluster orbit. A cross-section that decreases with velocity can therefore drive rapid evolution within the member's inner region while suppressing host-subhalo evaporation and changes to the cluster-scale halo \citep{LoebWeiner2011,Zeng2022,Zeng2025}. The cluster environment can then alter the collapse time in either direction. Tidal stripping removes part of the outer heat reservoir, and a concentrated stellar component deepens the central potential, both of which can accelerate collapse \citep{Nishikawa2020,YangYu2021,Zhong2023}. Tidal heating and host-subhalo evaporation instead add energy to the member and can delay the same evolution \citep{Zeng2022,Zeng2025}.

Velocity-dependent self-interactions are particularly relevant because they can enhance evolution within member galaxies while suppressing evaporation during high-velocity encounters \citep{LoebWeiner2011,Zeng2022, Zeng2025}. This is consistent with the empirical requirement that any successful explanation increases the projected mass within the central few kiloparsecs without strongly altering the outer tidal extent. Lensing-inferred truncation radii of observed cluster members remain broadly consistent with those of matched CDM subhalos \citep{Chiang2026}, while \citet{Natarajan+2026} similarly find that the discrepancy lies primarily in the inner rather than the outer mass distribution. As noted above, the observed and simulated members also have strongly discrepant radial distributions within the cluster, as highlighted previously and reiterated in recent work by \citet{Natarajan+2017,Natarajan+2026}. The problem is therefore not simply the survival of simulated cluster members, but where those members lie in the large-scale cluster potential and whether they retain sufficiently dense centers to produce the observed GGSL signal.

In this work, we separate the effects of member internal structure and the environment directly. We analyze the \ObservedClusterCount\ HST-imaged strong-lensing clusters, Abell~2744, Abell~S1063, Abell~370, MACS~J0416, MACS~J1206, and PSZ1~G311, together with ensembles of simulated analogs selected at comparable lens redshifts from the TNG-Cluster suite \citep{Nelson2024}. For every selected member, we compare its lensing strength in isolation, $\theta_{\rm E, iso}$, with that obtained after embedding it in the local cluster field, $\theta_{\rm E, cl}$. We separately identify members whose surrounding, member-excluded environmental field is already critical. 

\citet{King2007} showed that the convergence and shear supplied by a cluster can increase image separations, allowing less massive member galaxies to produce detectable strong-lensing events. Building on this picture, we ask whether observed clusters and cosmological simulations contain the same fraction of members that meet the GGSL criterion only when the local cluster field is included.

This decomposition leads to five mutually exclusive lensing classes. Members in cluster-critical environments are classified as \emph{Major Perturbers} if they satisfy the adopted strong lensing criterion in isolation, and as \emph{Minor Perturbers} otherwise. Members in cluster-subcritical environments are classified as \emph{Intrinsic Lenses} if they meet the adopted strong lensing criterion in isolation, \emph{Boosted Lenses} if they cross it only after the convergence of the local cluster field is included, and \emph{Inert Members} if they remain subcritical even with environmental assistance. This classification allows us to determine whether the GGSL discrepancy arises from differences in member internal lensing strength, differences in the environments occupied by those members, or both.

We then ask three related questions. First, do observed and simulated clusters place the same fraction of member galaxies in regions where the surrounding cluster field is already critical? Second, outside those regions, do galaxies of comparable mass have the same intrinsic and environmentally enhanced lensing strengths? Third, can a mass-conserving increase in central dark-matter density reproduce the observed member-lensing demographics without altering subhalo masses or tidal extents?

To disentangle these effects, we classify cluster members according to their intrinsic lensing strength and the contribution of their environment. As the GGSL discrepancy contains both a member's internal structure component and a spatial component, we demarcate these contributions in a unique way in this analysis by classifying cluster members into the distinct lensing types.

The paper is organized as follows. Section~\ref{sec:methods} describes the observed lens models, the selection of TNG-Cluster analogs, and the construction of simulated member profiles. In Section~\ref{sec:member_types}, we introduce the environmental decomposition and define the five member lensing classes. Section~\ref{sec:results} compares the observed and simulated populations, examines their dependence on mass, environment, and line-of-sight structure, and tests a phenomenological central-contraction model motivated by core-collapse SIDM for the inner density structure of subhalos. In Section~\ref{sec:discussion}, we discuss the implications for cluster subhalos, GGSL, and dark-matter physics. Throughout, we assume a flat $\Lambda$CDM cosmology with $H_0=70\ {\rm km\,s^{-1}\,Mpc^{-1}}$ and $\Omega_m=0.3$.

\section{Cluster Samples and Lens Models}
\label{sec:methods}

\subsection{Observed Strong-Lensing Clusters}
\label{sec:obs_cluster}

Our observed sample comprises \ObservedClusterCount\ massive strong-lensing clusters with parametric \Lenstool\ models: Abell~2744, Abell~S1063, MACS~J0416, MACS~J1206, PSZ1~G311.65$-$18.48, and Abell~370. These systems span the redshift range $z \simeq 0.31$--$0.44$, and we selected them for detailed study here because they combine high-quality HST imaging, extensive spectroscopic follow-up, and lens models that explicitly include resolved galaxy-scale perturbers. Together, they provide a well-studied observational baseline for comparing the lensing role of cluster members across different dynamical states and environments. The basic properties of the sample and the adopted model references are summarized in Table~\ref{tab:Cluster_Properties}.

\begin{table*}[t]
\centering
\caption{\textbf{Observed cluster sample.} The coordinates give the J2000
origin used to define relative positions in each adopted \Lenstool\ model.
The halo masses are the adopted literature $M_{200\mathrm{c}}$ values, in units of $10^{15}\,\Msun$. The quoted uncertainties correspond to $1\sigma$ where available.
\label{tab:Cluster_Properties}}
\footnotesize
\setlength{\tabcolsep}{5pt}
\begin{tabular}{lcccccc}
\hline\hline
Cluster & $z_{\rm cl}$ & R.A.$_0$ ($^\circ$) & Decl.$_0$ ($^\circ$) &
$M_{200\mathrm{c}}$ & Lens-model ref. & Mass ref. \\
\hline
Abell~2744  & 0.3072 &   3.586244 & $-30.400151$ & $2.06\pm0.42$
    & (1) & (5) \\
Abell~S1063 & 0.3475 & 342.183213 & $-44.530897$ & $1.98\pm0.60$
    & (2) & (6) \\
Abell~370 & 0.3750 & 39.969773 & $-1.572130$ & $2.20\pm0.41$
    & (10) & (12) \\
MACS~J0416  & 0.3972 &  64.038142 & $-24.067472$ & $1.14\pm0.27$
    & (3) & (6) \\
MACS~J1206  & 0.4398 & 181.550620 &  $-8.800936$ & $1.51\pm0.32$
    & (4) & (6) \\
PSZ1~G311 & 0.4436 & 237.52841667 & -78.19164472 & $0.85$
    & (9) & (11) \\
\hline
\end{tabular}
\par\medskip
\begin{minipage}{0.98\textwidth}
\raggedright
\textit{Note.} The redshifts of MACS~J0416 and MACS~J1206 are the
spectroscopic values from references (7) and (8), respectively. The
PSZ1~G311.65$-$18.48 value is the spectroscopic median from reference (9);
the other redshifts are given by their lens-model references.
Reference (6) reports masses in units of
$10^{14}\,h_{70}^{-1}\Msun$, with $h_{70}=1$ for the cosmology adopted here.
Reference (11) estimates its $M_{200\mathrm{c}}$ by converting the
Planck $M_{500}$ measurement and does not quote an uncertainty. Reference
(12) reports $M_{200\mathrm{c}}=(1.54\pm0.29)\times10^{15}\,
h^{-1}\Msun$; we use $h=0.7$ to obtain the value in the table.
\textbf{References:} (1) \citet{Bergamini2023:A2744};
(2) \citet{Beauchesne2023}; (3) \citet{Bergamini2023:M0416};
(4) \citet{Bergamini2019:M1206}; (5) \citet{Medezinski2016};
(6) \citet{Umetsu2018}; (7) \citet{Maier2016};
(8) \citet{Girardi2015}; (9) \citet{Pignataro2021};
(10) \citet{Gledhill2024:A370}; (11) \citet{Meneghetti2022};
(12) \citet{Umetsu2022:A370}.
\end{minipage}
\end{table*}

\textbf{Abell~2744} (A2744) is an HFF cluster at redshift $z=0.3072$ undergoing an active, multi-component merger, with several mass and gas substructures in and around its core \citep{Merten2011,Jauzac2016}. We adopt the \Lenstool\ model of \citet{Bergamini2023:A2744}, which combines HST imaging with MUSE spectroscopy and uses 90 spectroscopically confirmed multiple images and stellar velocity dispersions for 85 cluster members.

\textbf{Abell~S1063} (AS1063; also known as RXC~J2248.7$-$4431) is at redshift $z=0.3475$ and was observed by both CLASH and HFF \citep{Postman2012,Lotz2017}. We adopt the \Lenstool\ model of \citet{Beauchesne2023}, which jointly fits the strong-lensing constraints, the X-ray-emitting gas, and the dynamics of cluster members. The model treats the intra-cluster gas and collisionless matter as separate spatial components.

\textbf{Abell~370} (A370) is an HFF cluster at redshift $z=0.3750$ with a disturbed, bimodal core and evidence for elongation along the line-of-sight \citep{Umetsu2022:A370}. We adopt the CANUCS \Lenstool\ model of \citet{Gledhill2024:A370}, which adds JWST NIRCam and NIRISS observations to the existing HST imaging and MUSE spectroscopy.

\textbf{MACS~J0416.1$-$2403} (M0416) is an HFF cluster at redshift $z=0.3972$ with two central mass concentrations and an unrelaxed dynamical state \citep{Balestra2016}. We adopt the \Lenstool\ model of \citet{Bergamini2023:M0416}, constrained by 237 spectroscopically confirmed multiple images, stellar kinematics for 64 cluster members, and the X-ray-derived gas distribution.

\textbf{MACS~J1206.2$-$0847} (M1206) is at redshift $z=0.4398$ and was identified in the X-ray-selected MACS survey \citep{Ebeling2001}. It was later observed by CLASH and the CLASH-VLT spectroscopic program \citep{Postman2012,Rosati2014}. We adopt the \Lenstool\ model of \citet{Bergamini2019:M1206}, which combines spectroscopically confirmed multiple images with MUSE stellar velocity dispersions to constrain the scaling relations for the member-galaxy subhalos.

\textbf{PSZ1~G311.65$-$18.48} (PG311) is at redshift $z=0.4436$ and was selected for this comparison through its Planck Sunyaev-Zel'dovich signal rather than its strong-lensing properties \citep{Meneghetti2022}. It hosts the highly magnified Sunburst Arc \citep{Sharon2022:PG311}. We adopt the \Lenstool\ model of \citet{Pignataro2021}, which combines HST imaging and MUSE spectroscopy and is constrained by 62 multiple images from 17 families, as well as stellar velocity dispersions for 15 cluster members.

\subsection{Simulated Clusters}
TNG-Cluster extends the IllustrisTNG cosmological hydrodynamical simulations \citep{Pillepich2018, Nelson2019} to a large ($1\,{\rm Gpc}$) parent N-body simulation in which 352 cluster targets are identified, then re-simulated with higher resolution and with the IllustrisTNG galaxy formation model \citep{Nelson2024}. The mass resolution of the dark matter particles is $6.1\times 10^7\,{\rm M_\odot}$, and the mean baryonic cell mass is $1.2\times 10^7\,{\rm M_\odot}$, with a gravitational softening of $1.48\,{\rm kpc}$ for the dark matter and star particles, and a minimum adaptive softening for the gas of $370\,{\rm pc}$.

The features of the galaxy clusters in this suite have been studied extensively, such as the properties of the intra-cluster medium \citep{Lehle2024:tng_cluster_cool_cores, Truong2024:tng_cluster_xray_ICM}, radio relics resulting from mergers, and their ability to constrain self-interacting dark matter  \citep{Lee2024:tng_cluster_radio_relics, Jee2026:tng_cluster_radio_relics_sidm}, and the properties of cluster satellites \citep{Rohr2024:tng_cluster_cgm} and central galaxies \citep{Toptun2026:centrals}. 

\subsection{Summary of cluster lens modeling}\label{sec:lens_models}

In this section, we summarize the standard methodology employed to model the mass distribution of observed cluster lenses. This technique fundamentally relies on mapping the overall cluster mass using parametric density profiles, typically by decomposing the total mass into smooth, larger-scale dark matter halo components and smaller-scale perturbations, subhalos, associated with individual cluster member galaxies. We specifically favor this parametric approach because it provides a mathematical framework that is directly transferable to and easily comparable to simulated cluster analogs in cosmological simulations. This allows for a one-to-one comparison between observed subhalo populations and theoretical $\Lambda$CDM predictions. A comprehensive discussion of the mathematical formalism behind this procedure can be found in \citet{Natarajan1997,Natarajan2024}.

\subsubsection{Lensing formalism}

We briefly outline the lensing properties that are modeled with the mass distributions described above. The fundamental geometric framework of gravitational lensing connects the unlensed angular position of a background source, $\vec{\beta}$, to its observed angular position on the sky, $\vec{\theta}$. This mapping is described by the lens equation:
\begin{equation}
    \vec{\beta} = \vec{\theta} - \vec{\alpha}(\vec{\theta})
\end{equation}
where $\vec{\alpha}(\vec{\theta})$ is the scaled deflection angle. For a thin lens, the deflection angle can be expressed as the gradient of a two-dimensional, scaled projected lensing potential, $\psi(\vec{\theta})$, such that $\vec{\alpha} = \nabla\psi$. The lensing potential is intrinsically linked to the underlying mass distribution of the cluster through the Poisson equation, $\nabla^2\psi = 2\kappa(\vec{\theta})$. 

Here, $\kappa(\vec{\theta})$ represents the convergence, which is the dimensionless surface mass density of the lens, defined as the ratio of the projected surface mass density, $\Sigma(\vec{\theta})$, to the critical surface mass density, $\Sigma_{\mathrm{crit}}$ given by:
\begin{equation}
    \kappa(\vec{\theta}) = \frac{\Sigma(\vec{\theta})}{\Sigma_{\mathrm{crit}}}
\end{equation}
The critical density depends entirely on the cosmological distances involved in the lensing system, taking the form
\begin{equation}
    \Sigma_{\mathrm{crit}} = \frac{c^2}{4\pi G} \frac{D_s}{D_l D_{ls}}
\end{equation}
where $c$ is the speed of light, $G$ is the gravitational constant, and $D_l$, $D_s$, and $D_{ls}$ are the angular diameter distances to the lens, to the source, and between the lens and the source, respectively. 

To quantify the distortion and amplification of source images, we consider the local lens mapping, which is characterized by the Jacobian matrix, $\mathcal{A}$, of the lens equation:
\begin{equation}
    \mathcal{A}_{ij}
    = \frac{\partial \beta_i}{\partial \theta_j} = \delta_{ij} - H_{ij},
\end{equation}
where $H_{ij}\equiv\partial^2\psi/
(\partial\theta_i\partial\theta_j)$ are the components of the
lensing-potential Hessian, $i,j\in\{x,y\}$, and $\delta_{ij}$
is the Kronecker delta. The Jacobian is the inverse magnification
matrix and is conventionally expressed in terms of the convergence,
$\kappa$, and the two shear components, $\gamma_1$ and $\gamma_2$:
\begin{equation}
    \mathcal{A} = 
    \begin{pmatrix} 
    1 - \kappa - \gamma_1 & -\gamma_2 \\ 
    -\gamma_2 & 1 - \kappa + \gamma_1 
    \end{pmatrix}
\end{equation}
While the convergence $\kappa$ governs the isotropic magnification of the source, the shear $\gamma = \sqrt{\gamma_1^2 + \gamma_2^2}$ induces anisotropic stretching. The total magnification of an image, $\mu$, is given by the inverse of the determinant of the Jacobian:
\begin{equation}
    \mu = \frac{1}{\det \mathcal{A}} = \frac{1}{(1-\kappa)^2 - \gamma^2}
\end{equation}

In the strong lensing regime, critical curves result where the lens mapping becomes singular, $\det\mathcal{A}=(1-\kappa)^2-\gamma^2=0$. A critical curve can therefore
occur at $\kappa<1$ when the shear is nonzero. These regions define the critical curves in the lens plane, which map to caustics in the source plane. When a background source crosses a caustic, its magnification theoretically approaches infinity, leading to the formation of multiple highly distorted images, giant arcs, and Einstein rings that are seen in HST \& JWST images such as those described in Sec.~\ref{sec:obs_cluster}.

\subsubsection{Parametric Mass Modeling with \texttt{Lenstool}}

The publicly available software \texttt{Lenstool} \citep{Jullo2007} constructs a parametric model of the cluster mass distribution by decomposing the total projected gravitational potential, $\psi_{\mathrm{tot}}$, into a combination of large-scale smooth dark matter halos and small-scale perturbations associated with individual cluster member galaxies. Mathematically, this decomposition is expressed as proposed by the conceptual model presented in \citet{Natarajan1997,Jullo2007}:
\begin{align}
    \psi_{\rm tot}
    =
    \sum_i \psi_i^{\rm halo}
    +
    \sum_j \psi_j^{\rm subhalo}
    +
    \psi_{\rm ext},
\end{align}
where $\psi^\text{halo}_i$ represents Mpc-scale halos associated with the smooth large-scale cluster gravitational potentials; $\psi^\text{subhalo}_j$ are kpc-scale subhalos associated with cluster member galaxies, and $\psi_{\rm ext}$ the gravitational potential of a constant external shear field.

To constrain this model, a robust observational dataset is required. First, cluster members are identified primarily via their photometric properties (e.g., residing on the cluster red sequence) and, whenever possible and preferably, confirmed via spectroscopy. Second, secure sets of strongly lensed multiple-image families are cataloged. Secure source redshifts fix the relevant distance ratios, and multiple measured source redshifts ($z_{\mathrm{spec}}$) help break the mass-sheet degeneracy and provide an absolute calibration for the cluster mass.

Because modeling hundreds of individual galaxies with independent free parameters would render the system highly computationally prohibitive and degenerate, \texttt{Lenstool} reduces the parameter space by tying the properties of the galaxy-scale subhalos to their observed luminosities. Each cluster member is typically modeled using a dual Pseudo Isothermal Elliptical (dPIE) mass profile \citep{Eliasdottir2007}, characterized by a central velocity dispersion $\sigma_0$, a core radius $r_{\mathrm{core}}$, and a truncation or cut radius $r_{\mathrm{cut}}$. These parameters are scaled relative to a characteristic cluster galaxy with luminosity $L^\star$, following established empirical scaling relations, i.e. the Faber-Jackson relation and the Fundamental Plane:
\begin{eqnarray}
    \sigma_0 &=& \sigma_0^\star \left( \frac{L}{L^\star} \right)^{1/4} \\
    r_{\mathrm{core}} &=& r_{\mathrm{core}}^\star \left( \frac{L}{L^\star} \right)^{1/2} \\
    r_{\mathrm{cut}} &=& r_{\mathrm{cut}}^\star \left( \frac{L}{L^\star} \right)^{\alpha}
\end{eqnarray}
where $\sigma_0^\star$, $r_{\mathrm{core}}^\star$, and $r_{\mathrm{cut}}^\star$ are the parameters of the $L^\star$ galaxy, and $\alpha$ is typically fixed to $0.5$ for a constant mass-to-light ratio scaling. By enforcing these relations, the individual subhalo potentials are fully described by the global parameters of the $L^\star$ galaxy, significantly reducing the number of free parameters.

With the galaxy-scale parameters tied to the scaling relations, the remaining free parameters, primarily those governing the position, ellipticity, position angle, and mass limits of the large-scale halos, as well as the $L^\star$ scaling parameters, are optimized. \texttt{Lenstool} performs this optimization using a Bayesian Markov Chain Monte Carlo (MCMC) sampler. The MCMC algorithm explores the multidimensional parameter space to minimize the root-mean-square (rms) scatter between the observed positions of the multiple images $\vec{\theta}_{\mathrm{obs}}$, and those predicted by the lens equation $\vec{\theta}_{\mathrm{pred}}$. 

For each cluster member, we extract the best-fit position, ellipticity, orientation, velocity dispersion, and truncation radius. We interpret the truncation parameter $r_{\rm cut}$ as a proxy for the tidal radius, following previous lensing-based tidal radius analyses \citep{Natarajan2002, Natarajan2009, Limousin2007, Sifon2018, Chiang2026}. We then map the \texttt{Lenstool} parametrization into equivalent \texttt{Lenstronomy} components \citep{Birrer2018} so that observations and simulations can be analyzed with a common lensing pipeline. For observed members, we define $M_{\rm sub}$ as the projected mass enclosed within $r_{\rm cut}$.

\subsection{Selection of TNG-Cluster analog ensembles}
\label{subsec:tng_selection}

As simulation counterparts, we draw from the TNG-Cluster simulation suite \citep{Nelson2024}, a set of high-resolution MHD zoom-in simulations of massive clusters utilizing the IllustrisTNG galaxy formation model \citep{Pillepich2018, Nelson2019}. 
For each observed cluster, we consider TNG-Cluster halos at the nearest available redshift with $M_{200\mathrm{c}}>10^{15}\,M_\odot$. Following \citet{Meneghetti2020}, we characterize the primary tangential critical curve by its Einstein radius, $\theta_{\rm E,p}=\sqrt{A_{\rm p}/\pi}$, where $A_{\rm p}$ is the angular area enclosed by the cluster-scale tangential critical curve. We choose as the projection axis for each simulated cluster the one that maximizes $\theta_{\rm E,p}$ among 100 randomly sampled directions, uniform on the sphere. For each observed cluster, we retain the ten simulated clusters at the same redshift that most closely match in $\theta_{\rm E,p}$, yielding 60 simulated analogs. This procedure can select the same simulated object, viewed at different times, as a match for two or more observed objects. This could introduce correlations in the combined simulated dataset that under-predict scatter in some quantities of interest, so in any case where this may occur, we will also consider single observed objects and their matched counterparts in isolation (cf. Appendix~\ref{appendix:cluster_fractions} Table~\ref{tab:cluster_fractions}).
We select the brightest members in each simulated analog by their rest-frame $V$-band luminosity, retaining the same number of members as in the corresponding observed lens-model catalog.
For simulated members, we define $M_{\rm sub}$ as the total gravitationally bound mass reported by \textsc{Subfind}, including dark matter, stars, gas, and black holes.

The particles in one simulated cluster used to compute lensing quantities here are those gravitationally bound to the cluster, as determined by a friends-of-friends algorithm. This constitutes a single massive group in the TNG-Cluster group catalog. Notably, this does not include correlated structure along the line of sight, which contributes to the lensing strength of the cluster and whose galaxies may be erroneously identified as cluster members in observations. When computing the lensing potential and its derivatives due to the simulated central halo, we utilize a $400\arcsec$ field of view to avoid boundary truncation effects \citep{Vyvere2020, Roche2026}, sampled on a $4096\times4096$ grid with $0.0977\arcsec$ per pixel. 

\subsection{Mass Profiles of Simulated Clusters \& their members}
\label{subsec:profile_fits}


We construct the lensing fields of the simulated analogs by smoothing the particles of the main halo with analytic models for its satellite subhalos. For the main halo, we smooth and project the dark-matter, gas, stellar, and black-hole particles bound to the central halo using \texttt{py-sphviewer2}\footnote{\url{https://github.com/alejandrobll/py-sphviewer2}} \citep{Benitez2025pysphviewer2}. 
We model the satellites separately using analytical profiles.

For each satellite, we model the dark matter with a spherical truncated NFW (tNFW) profile \citep{BaltzMarshallOguri2009tNFW},
\begin{align}
\rho_{\rm NFW}(r) &= \frac{\rho_{0,\rm DM}} {(r/r_{\rm s,DM})(1+r/r_{\rm s,DM})^2}, \\
\rho_{\rm tNFW}(r) &= \rho_{\rm NFW}(r) \frac{r_{\rm t,DM}^2}{r^2+r_{\rm t,DM}^2},
\end{align}
with free parameters $(\rho_{0,\rm DM},r_{\rm s,DM},r_{\rm t,DM})$.

We model the bound gas with a spherical pseudo-Jaffe profile,
\begin{equation}
\rho_{\rm PJ}(r)
= \frac{\rho_{0,\rm gas}}
{\left[1+(r/r_{\rm a})^2\right]
 \left[1+(r/r_{\rm s,gas})^2\right]},
\end{equation}
where the normalization, core radius $r_{\rm a}$, and outer scale radius $r_{\rm s,gas}$ are fitted independently for each satellite.

We fit each component with at least 250 bound particles or cells in ten logarithmically spaced projected annuli, comparing the measured and model-predicted masses. We use Poisson statistics for dark matter and a scaled-Poisson approximation for gas, accounting for unequal cell masses through $s_j=\sum_{i\in j}m_i^2/\sum_{i\in j}m_i$.

Each satellite's stellar component is modeled with a Hernquist profile \citep{Hernquist1990profile}, fixing its total mass $M_\star$ and scale radius $a=r_{1/2}/(1+\sqrt{2})$ from the catalogued bound stellar mass and three-dimensional stellar half-mass radius. Black holes are represented as point masses.

We combine the particle-based main-halo lensing field with the analytic satellite components in \texttt{lenstronomy}, allowing us to calculate lensing observables for simulated and observed clusters within the same framework. Modeling the components separately retains their different radial structures without imposing a single profile on the total satellite mass.

\section{Environmental Regimes and Member Lensing Classes}
\label{sec:member_types}

Our classification is hierarchical. We first assign each member to one of two environmental regimes according to whether the field remaining after removing the target member is critical at its position; this field includes the smooth cluster components and all other members, excluding only the target member. Members in the cluster-critical regime perturb an existing cluster-scale critical curve and are classified as Major or Minor Perturbers. Members in the cluster-subcritical regime can generate a distinct secondary critical curve and are classified as Intrinsic Lenses, Boosted Lenses, or Inert Members. We refer to these five mutually exclusive outcomes collectively as the \emph{lensing classes}, defined in Sec.~\ref{subsec:member_types}. The five classes describe the lensing role of every selected cluster member, while only Intrinsic and Boosted Lenses are direct producers of distinct GGSL secondary critical curves in cluster-subcritical regions. Major and Minor Perturbers modify only the cluster-scale critical structure.

As noted above, the GGSL cross section of a cluster depends not only on the internal structure of its member galaxies but also on how those galaxies are embedded in the local cluster tidal field. A subhalo that is subcritical in isolation can become a strong lens if placed sufficiently close to the critical region of the host cluster \citep{BlandfordKochanek1987, SchneiderEhlersFalco1992, Meneghetti2003}. We separate these contributions by measuring the lensing strength of each member in isolation and after including the boost provided by the local cluster environment. This comparison determines whether a member lenses on its own, becomes a strong lens because of the cluster environment, or is proximate to an existing cluster critical curve.

We compute all lensing quantities for sources at $z_{\rm s}=4$. For each member $k$, we define its environment by removing that member from the full cluster model while keeping every other mass component fixed:
\begin{equation}
\psi_{{\rm env},k}=\psi_{\rm full}-\psi_k.
\end{equation}
The environment, therefore, includes the smooth cluster components and all other selected members. When line-of-sight structure is included, we add its local convergence and shear to this environmental field.

\subsection{Local lensing strength decomposition}
\label{subsec:einstein_radii}

\begin{figure*}[ht]
    \centering
    \includegraphics[width=\linewidth]{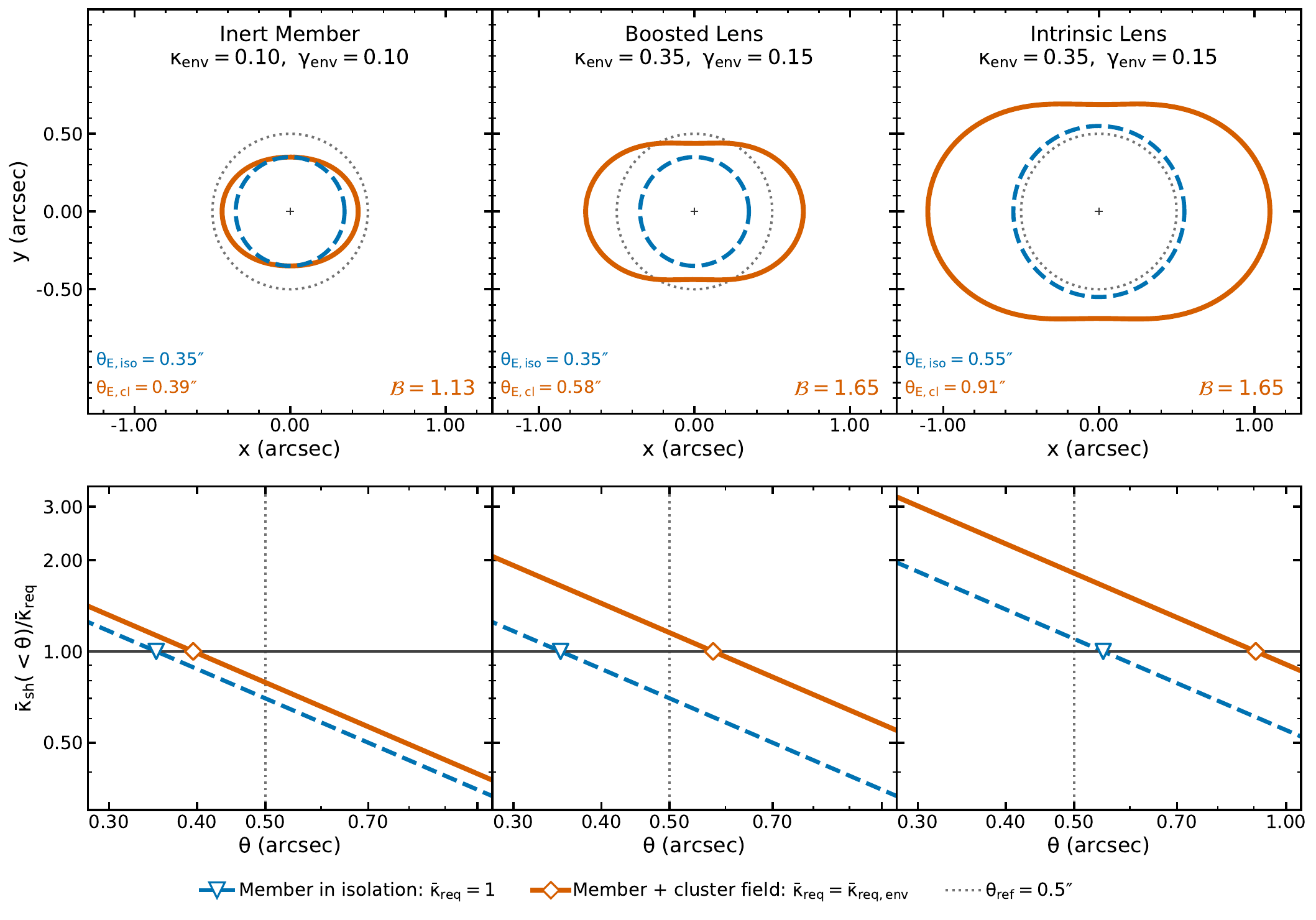}
\caption{\textbf{Environmental impact on member lensing strength for an Inert Member, a Boosted Lens, and an Intrinsic Lens:} each is modeled as a singular isothermal sphere in a uniform, subcritical external convergence and shear field, for illustration. Top panels show the tangential critical curves in isolation (blue dashed) and with the cluster environment included (orange solid). The environmental shear distorts the otherwise circular critical curves. Gray dotted circles mark the reference radius $\theta_{\rm ref}=0\farcs5$. Bottom panels show the member's mean enclosed convergence normalized by the required value to strongly lens, $\bar{\kappa}_{\rm req}=1$ in isolation and $\bar{\kappa}_{\rm req}=\bar{\kappa}_{\rm req,env}$ in the external field. Their intersections with unity give the corresponding Einstein radii, marked by triangles and diamonds. The first two members have identical isolated profiles, but the stronger environment in the middle column raises $\theta_{\rm E,cl}$ above the reference radius. The rightmost member is critical even in isolation. The annotated boost is $\mathcal{B}=\theta_{\rm E,cl}/\theta_{\rm E,iso}$.}
    \label{fig:environmental_assistance_schematic}
\end{figure*}

Galaxy–galaxy strong lensing depends on the subhalo’s own mass distribution and can be enhanced by the convergence and shear of its surroundings. Because convergence, shear components, and deflection angles are additive within the cluster plane, we can separate the member’s contribution from that of its environment. For each member, we consider four lensing configurations:
\begin{enumerate}
    \item an \emph{environment} model containing the full cluster lens with only the target member removed;
    \item an \emph{isolated subhalo} model containing only the galaxy-scale perturber of interest, evaluated at the same sky position but without the underlying smooth cluster environment;
    \item a \emph{cluster-field member} model in which the same perturber is embedded in a uniform external convergence and shear field equal to the locally measured values from the environment model; and
    \item the \emph{full cluster} model, containing the smooth cluster components and all selected members, which we use to describe the cluster-scale critical structure.
\end{enumerate}

Together, these configurations allow us to separate the intrinsic lensing strength of each member from the contribution provided by the cluster environment. The isolated-member and cluster-field models define $\theta_{\rm E,iso}$ and $\theta_{\rm E,cl}$, respectively. The local environmental convergence and shear, $(\kappa_{\rm env}, \gamma_{\rm env})$, describe the combined field of the smooth cluster components and the other selected members. We derive these quantities from the Hessian of $\psi_{{\rm env},k}$ averaged over a $0\farcs2$ window centered on member $k$:

\begin{align}
\kappa_{\rm env}
&\equiv \frac{1}{2}
\left\langle
H_{xx,{\rm env},k}+H_{yy,{\rm env},k}
\right\rangle, \\
\gamma_{1,\rm env}
&\equiv \frac{1}{2}
\left\langle
H_{xx,{\rm env},k}-H_{yy,{\rm env},k}
\right\rangle, \\
\gamma_{2,\rm env}
&\equiv
\left\langle H_{xy,{\rm env},k}\right\rangle, \\
\gamma_{\rm env}
&\equiv
\left(\gamma_{1,\rm env}^2+\gamma_{2,\rm env}^2\right)^{1/2},
\end{align}
where
$H_{ij,{\rm env},k}\equiv
\partial^2\psi_{{\rm env},k}/
(\partial\theta_i\partial\theta_j)$,
with $i,j\in\{x,y\}$, are the components of the target-excluded
environmental Hessian for member $k$.
The brackets $\langle\cdot\rangle$ denote the spatial average
over the window centered on that member.
We denote the resulting constant environmental Hessian by
$\overline{\mathbf{H}}_{{\rm env},k}
\equiv\langle\mathbf{H}_{{\rm env},k}\rangle$.
For brevity, we suppress the member index on the environmental
convergence and shear.
Averaging the Hessian components reduces small-scale numerical fluctuations while retaining the cluster field that can enhance the lensing efficiency of a member galaxy \citep{BlandfordKochanek1987, SchneiderEhlersFalco1992, Meneghetti2003, Meneghetti2020}.

The local cluster field can also be written in terms of the tangential and radial eigenvalues,
\begin{align}
\lambda_{t,\rm env} &=1-\kappa_{\rm env}-\gamma_{\rm env}, \\
\lambda_{r,\rm env} &=1-\kappa_{\rm env}+\gamma_{\rm env}.
\end{align}

When $\lambda_{t,\mathrm{env}}\leq 0$, the member-excluded environmental field is already tangentially critical. In this case, the member modifies the existing cluster critical curve rather than producing a separate member-scale curve. When $\lambda_{t,\rm env}>0$, the member must provide the remaining convergence needed to form a critical curve.

In the standard GGSL treatment, member-scale Einstein radii are measured from the area enclosed by a secondary critical curve. Near the cluster critical curve, however, the two can merge, making it difficult to assign a separate curve to an individual member. To measure each member’s lensing strength without requiring this association, we use its mean enclosed convergence:

\begin{equation}
\bar{\kappa}_{\rm sh}(<\theta) = \frac{1}{\pi\theta^2}
\int_{|\boldsymbol{\theta}'|<\theta} \kappa_{\rm sh}(\boldsymbol{\theta}')\,
{\rm d}^2\boldsymbol{\theta}' = \frac{ \left\langle\alpha_{r,\rm sh}(\theta)\right\rangle }{\theta},
\label{eq:kbar_sh}
\end{equation}

where $\alpha_{r,\rm sh}$ is the radial deflection produced by the member, averaged around a circle of radius $\theta$. The second equality follows from
$\boldsymbol{\nabla}\cdot\boldsymbol{\alpha}=2\kappa$
and the two-dimensional divergence theorem, and therefore does not require axial symmetry \citep[see Chapters~5 and 6]{SchneiderEhlersFalco1992}. It allows us to measure the member's lensing strength without identifying a separate closed critical curve.

For each member, we calculate $\bar{\kappa}_{\rm sh}(<\theta)$ as a function of angular distance from its center. Following \citet{Meneghetti2020}, we adopt $\theta_{\rm ref}=0\farcs5$ as the minimum Einstein radius in the GGSL sample. This lower limit excludes galaxy-scale critical curves near the minimum angular scale that can be measured reliably in HST imaging, and ensures only well-converged objects are used in the simulated data.

The cluster convergence and shear reduce the mean convergence that a member must supply to become critical. For a singular isothermal sphere in a uniform, subcritical external field, the required mean convergence is
\begin{equation}
\bar{\kappa}_{\rm req,env} \equiv
\frac{(1-\kappa_{\rm env})^2-\gamma_{\rm env}^2}
{\left[(1-\kappa_{\rm env})^2+\gamma_{\rm env}^2/2\right]^{1/2}}.
\label{eq:kappa_req_env}
\end{equation}
Without shear, this reduces to
$\bar{\kappa}_{\rm req,env}=1-\kappa_{\rm env}$.

We compare this required value with each member's convergence profile to define the isolated and cluster-field radii through the outermost solutions of
\begin{align}
    \bar{\kappa}_{\rm sh}(<\theta_{\rm E,iso}) &= 1, \\
    \bar{\kappa}_{\rm sh}(<\theta_{\rm E,cl})
    &= \bar{\kappa}_{\rm req,env}.
\end{align}
We define the isolated radius for every member and the cluster-field radius only where $\lambda_{t,\rm env}>0$. Together, these radii separate the member's own lensing strength from the contribution of its surroundings, as illustrated in Figure~\ref{fig:environmental_assistance_schematic}.

Our treatment of the surrounding environment is exact for a singular isothermal
sphere and provides a useful approximation for other member profiles, with $\bar{\kappa}_{\rm req,env}$ offering a simple measure of the lensing assistance from the environment. Appendix~\ref{appendix:hessian} shows close agreement with radii measured from critical curves located using the Hessian.

Following the picture developed by \citet{King2007}, we quantify how the surrounding field increases a member's lensing scale. For members with $\theta_{\rm E,iso}>0$ and a defined cluster-field radius, we define the environmental boost as
\begin{equation}
\mathcal{B}\equiv
\frac{\theta_{\rm E,cl}}{\theta_{\rm E,iso}}.
\label{eq:environmental_boost}
\end{equation}
King measured the enhancement using the semimajor axis of the critical curve, whereas we use its effective Einstein radius. For an isothermal member, $\mathcal{B}=1/\bar{\kappa}_{\rm req,env}$. More generally, the boost also depends on the member's radial mass profile because $\theta_{\rm E,iso}$ and $\theta_{\rm E,cl}$ are obtained directly from $\bar{\kappa}_{\rm sh}(<\theta)$.

\subsection{Classifying Members by their Lensing Class}
\label{subsec:member_types}

\begin{figure*}[ht!]
    \centering
    \includegraphics[width=\linewidth]{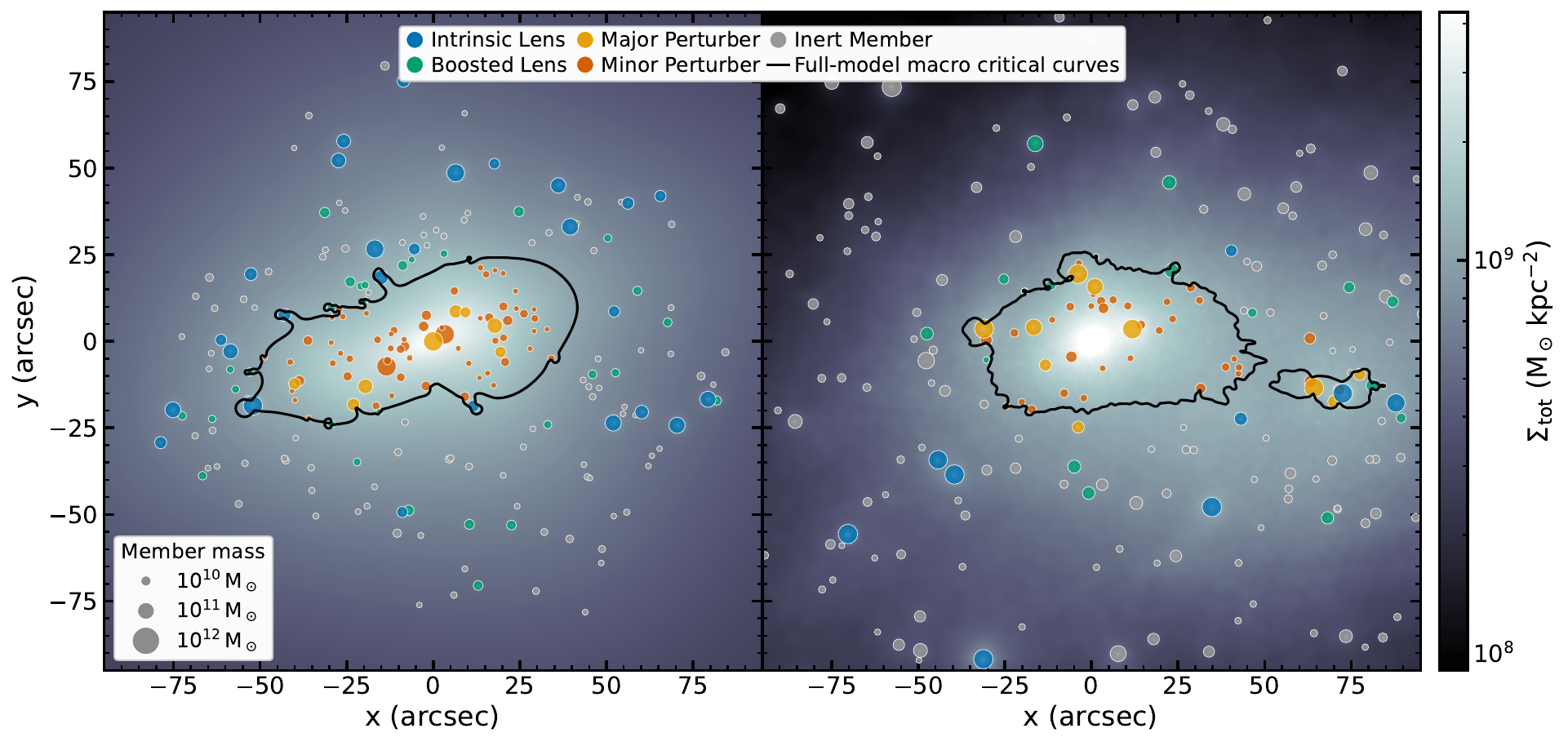}
\caption{\textbf{Observed and simulated member classes in MACS J1206:} the left panel shows the observed cluster members, while the right panel shows its TNG-Cluster simulated analog. Colors indicate the five lensing classes, and marker area scales with member mass. The right panel also shows the projected surface density of the particles bound to the simulated subhalos, including dark matter, gas, stars, and black holes. The black lines show the cluster-scale tangential critical curves of the full lens model for a source at $z_{\rm s}=4$; these curves provide geometric context and do not define the member classes.}
    \label{fig:subhalos}
\end{figure*}

We assign the five lensing classes using three quantities: $\lambda_{t,\rm env}$, $\theta_{\rm E,iso}$, and $\theta_{\rm E,cl}$. We first separate members according to whether their environment is critical or supercritical ($\lambda_{t,\rm env}\leq0$). These members are classified as Perturbers. For the remaining members, the isolated and cluster-field Einstein radii determine whether they strongly lens.



We assign each cluster member to one of five lensing classes using the single reference scale $\theta_{\rm ref}=0\farcs5$, although adopting $1''$ leads to the same main conclusions (Appendix~\ref{appendix:reference_radius}).
Members on the cluster-scale critical structure are divided into two types:

\begin{itemize}
    \item \emph{Major Perturbers} have
    $\theta_{\rm E,iso}\geq\theta_{\rm ref}$. These members are strong lenses on their own, but at their cluster positions, their critical curves merge with the larger cluster curve.

    \item \emph{Minor Perturbers} have
    $\theta_{\rm E,iso}<\theta_{\rm ref}$. They do not meet the GGSL threshold in isolation but can still distort the cluster critical curve \citep{Meneghetti2000}.
\end{itemize}

We divide the remaining members, which lie outside the cluster-scale critical structure and are contributors to GGSL, according to their isolated and cluster-field Einstein radii:

\begin{itemize}
    \item \emph{Intrinsic Lenses} have
    $\theta_{\rm E,iso}\geq\theta_{\rm ref}$. They meet the GGSL threshold without assistance from the cluster environment.

    \item \emph{Boosted Lenses} have
    $\theta_{\rm E,iso}<\theta_{\rm ref}\leq\theta_{\rm E,cl}$. They fall below the threshold in isolation but cross it after the local cluster convergence and shear are included.

    \item \emph{Inert Members} have
    $\theta_{\rm E,cl}<\theta_{\rm ref}$. They remain below the GGSL threshold even after the inclusion of the cluster environment.
\end{itemize}
Figure~\ref{fig:subhalos} shows the spatial distribution of these five classes in MACS~J1206 and its simulated analog.

These lens types therefore also describe how the inner structure of a member changes its role in GGSL. A member with a dense inner mass distribution can meet the lensing threshold in isolation, with the cluster environment providing only an additional boost. A member slightly below the threshold can become a strong lens when embedded in the cluster field. By contrast, a member that has lost mass through tidal stripping or heating can remain below the threshold even with the aid of the cluster environment. In simulations, numerical over-merging or artificial disruption can produce a similar loss of substructures \citep{Merritt1983, Moore1996, Ghigna1998, Gnedin1999, Limousin2007, vanDenBoschOgiya2018, Errani2020, Errani2021}.

Member mass remains a continuous property and does not define any of the class boundaries. The classification depends only on lensing quantities measured in the same way for the observed clusters and simulated analogs.

\subsection{The effect of line-of-sight structure}

In observations of galaxy cluster strong lenses, lensing models commonly neglect line-of-sight galaxies or approximate their mass contributions at the cluster redshift \citep{Raney2020:los_lensing}. Some models explicitly include a small number of significant foreground and background perturbers in the field of view \citep{Bergamini2023:M0416,Acebron2025:J0138,Betts2026:SLICE_SPTJ1150}, while other studies use multi-plane lensing to account for their different redshifts \citep{Chirivi2018:J0416,Raney2020:los_lensing}.
However, a significant cosmological volume remains unmodeled, with some effects such as $\sim$arcminute-scale deflections due to large-scale structure not being directly observable \citep{Seljak1994:LSS_lensing}, while other effects, such as perturbations due to interloping objects, are capable of producing changes in the \textit{relative} separation of lensed images by several arcseconds \citep{Host2012:los_lensing, Raney2020:los_lensing}. These effects will be subtly present in the observed data but not in the simulated data, which includes only the mass associated with the cluster object. It is therefore useful to quantify the effect of this unmodeled structure in the simulations.

The observed clusters are selected for presenting significant lensing evidence, and are thus likely to lie on over-dense lines of sight \citep{Bayliss2014}. To coarsely estimate the effect of a slightly over-dense line of sight on member classification, we repeat the classification of the simulation cluster members, including a small number of mass planes with mean density larger than the background density of the universe. These planes are extracted from light cones generated using the TNG300-1 hydrodynamical simulation box \citep{Nelson2019, Pillepich2018}, and the creation of these light cones is outlined in \citet{Roche2026}. Since TNG-Cluster comprises a series of zoom-in simulations of cluster-scale targets, a uniform box is preferable for generating line-of-sight structure planes. Note that TNG300 and TNG-Cluster use the same galaxy formation model. The chosen planes are drawn from the light cone constructed for group 0 in \citet{Roche2026}, chosen to be at redshifts $z=0.244$, $0.348$, $0.887$, $1.206$, and $1.604$. For simulated clusters matched in redshift to PSZ1 G311, MACS J1206, MACS J0416, and Abell 370, three of these planes are in the background of the cluster, with 2 in the foreground, whereas for those matched to Abell 2744 there are four planes in the background and one in the foreground. Abell S1063 is at approximately the same redshift as one of the chosen planes, in which case there are three background, one foreground, and one ``correlated structure''-like plane. The planes each have a $500\arcsec$ field of view, and contrast masses relative to the background density of $8.51\times 10^{12}\,\Msun$, $2.75\times 10^{13}\,\Msun$, $1.11\times 10^{14}\,\Msun$, $4.59\times 10^{14}\,\Msun$, and $3.10\times 10^{14}\,\Msun$, respectively. The total added mass in the cone above the mean density is then $9.16\times 10^{14}\,\Msun$, which is less mass than is present in the line of sight planes of, for example, the group 2 light cone in \citet{Roche2026} ($1.52\times 10^{15}\,\Msun$), thus representing an over-dense but reasonable line of sight. This is intended as an initial estimate of the role played by an over-dense line of sight, with a full light cone treatment left for future work.

We convert the planes' surface-density contrasts to convergence for sources at $z_{\rm s}=4$. We smooth each convergence map with a Gaussian of width $\sigma=10''$ and calculate the corresponding shear, so that the added structure acts as a smooth external field. We use the lens equation to sample each background plane at the ray position after deflection by the cluster. We approximate the smoothed line-of-sight structure as a weak external perturbation. For each cluster member, we evaluate how this perturbation changes the surrounding lensing field. To first order in this perturbation, and neglecting lens--lens coupling, the combined convergence and shear components are:
\begin{equation}
\begin{aligned}
\kappa_{\rm env}    &= \kappa_{\rm cl} + \kappa_{\rm LoS}, \\
\gamma_{i,\rm env}  &= \gamma_{i,\rm cl} + \gamma_{i,\rm LoS},
\qquad i = 1,2,
\end{aligned}
\end{equation}
where $\kappa_{\rm cl}$ and $\gamma_{i,\rm cl}$ model the field produced by the cluster environment, excluding the member whose lensing strength we are evaluating. We use the combined fields to recompute the member's embedded lensing strength and classification, keeping its mass profile fixed. We reuse the same five planes for all simulated analogs.

We treat the line-of-sight contribution as a local external field rather than performing a full multi-plane lensing calculation. The $10''$ smoothing scale is larger than the $0\farcs5$ member-lensing threshold, which motivates approximating the added convergence and shear as uniform across the member's lensing region. Sampling the planes along the cluster-deflected ray accounts for the change in where the ray crosses the background structure, but neglects the full coupling between lens planes. This approximation allows us to estimate how the added field changes member lensing strength while keeping the member profiles fixed.
\section{Results}
\label{sec:results}

We now present the results of our analysis, which separates differences in member internal structure from differences in spatial location by evaluating the lensing strength of observed clusters and their TNG-Cluster analogs. We then examine how these differences depend on member mass and cluster-centric radius, test whether line-of-sight structure changes the comparison, and finally determine whether a mass-conserving redistribution of existing dark matter toward the center is sufficient to recover much of the conditional lensing-class demographic while leaving the occupancy problem unresolved.

\subsection{Separating the cluster environment from intrinsic member lensing strength}
\label{subsec:results_intrinsic_member_strength}

\begin{figure*}[ht]
    \centering
    \includegraphics[width=\linewidth]{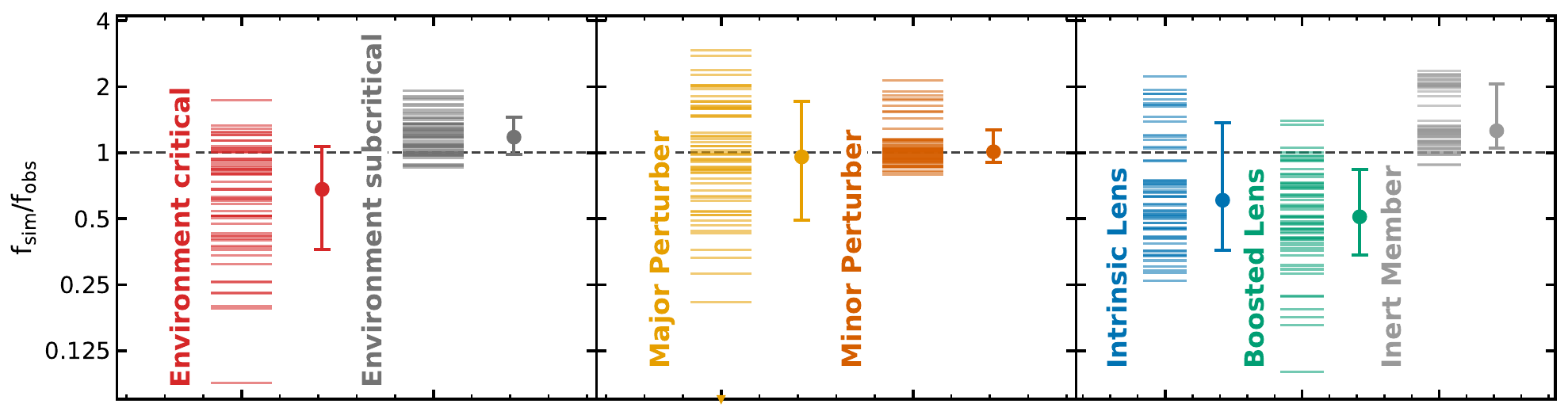}
\caption{
\textbf{Cluster environment and lensing-class demographics:} each short horizontal mark shows the fraction in one simulated analog divided by the fraction in its matched observed cluster. The dashed line at unity indicates agreement. The left panel indicates the fractions of members where the cluster environment is already critical. The middle panel shows the Major and Minor Perturber fractions among members in the cluster-critical region, while the right panel shows the Intrinsic Lens, Boosted Lens, and Inert Member fractions among members in cluster-subcritical regions. Filled circles and vertical intervals give the cluster-balanced median and 16th--84th percentiles across the 60 analogs. Downward triangles mark ratios below the plotted range.
\label{fig:ratio_barcode}}
\end{figure*}

Table~\ref{tab:type_fractions} summarizes the lensing-class fractions for bright members in the observed cluster sample and in the stacked TNG analog ensembles within the strong-lensing footprint, $R\lesssim 800$~kpc.
\begin{table*}[t]
    \centering
    \caption{Cluster environment and conditional lensing-class fractions for
    bright members in the \ObservedClusterCount\ observed clusters and
    \SimulatedWithoutLineOfSightAnalogCount\ matched TNG-Cluster analogs: the first block gives the fraction of all selected members in each
    environment. The remaining blocks give fractions normalized within the
    indicated environment. All entries are cluster-balanced mean fractions.
    The baseline simulated fractions exclude line-of-sight structure. The LoS
    column includes the adopted line-of-sight planes, while the cSIDM column
    applies the mass-conserving central contraction without line-of-sight
    structure.}
    \label{tab:type_fractions}
    \setlength{\tabcolsep}{5pt}
    \begin{tabular}{lcccc}
        \hline\hline
        Population & Obs. & Sim. & Sim. + LoS & Sim. + cSIDM \\
        \hline
        \multicolumn{5}{l}{\emph{Fraction of all selected members}} \\
        \quad Perturbers (cluster critical)
            & $\ObservedClusterCriticalRegionClusterBalancedMeanPercent\%$
            & $\SimulatedWithoutLineOfSightClusterCriticalRegionClusterBalancedMeanPercent\%$
            & $\SimulatedWithLineOfSightClusterCriticalRegionClusterBalancedMeanPercent\%$
            & $\SimulatedWithCoreCollapsedSIDMClusterCriticalRegionClusterBalancedMeanPercent\%$ \\
        \quad Nonperturbers (cluster subcritical)
            & $\ObservedClusterNoncriticalRegionClusterBalancedMeanPercent\%$
            & $\SimulatedWithoutLineOfSightClusterNoncriticalRegionClusterBalancedMeanPercent\%$
            & $\SimulatedWithLineOfSightClusterNoncriticalRegionClusterBalancedMeanPercent\%$
            & $\SimulatedWithCoreCollapsedSIDMClusterNoncriticalRegionClusterBalancedMeanPercent\%$ \\
        \hline
        \multicolumn{5}{l}{\emph{Within the cluster-critical region}} \\
        \quad Major Perturber
            & $\ObservedMajorPerturberWithinClusterCriticalRegionClusterBalancedMeanPercent\%$
            & $\SimulatedWithoutLineOfSightMajorPerturberWithinClusterCriticalRegionClusterBalancedMeanPercent\%$
            & $\SimulatedWithLineOfSightMajorPerturberWithinClusterCriticalRegionClusterBalancedMeanPercent\%$
            & $\SimulatedWithCoreCollapsedSIDMMajorPerturberWithinClusterCriticalRegionClusterBalancedMeanPercent\%$ \\
        \quad Minor Perturber
            & $\ObservedMinorPerturberWithinClusterCriticalRegionClusterBalancedMeanPercent\%$
            & $\SimulatedWithoutLineOfSightMinorPerturberWithinClusterCriticalRegionClusterBalancedMeanPercent\%$
            & $\SimulatedWithLineOfSightMinorPerturberWithinClusterCriticalRegionClusterBalancedMeanPercent\%$
            & $\SimulatedWithCoreCollapsedSIDMMinorPerturberWithinClusterCriticalRegionClusterBalancedMeanPercent\%$ \\
        \hline
        \multicolumn{5}{l}{\emph{Outside the cluster-critical region}} \\
        \quad Intrinsic Lens
            & $\ObservedIntrinsicLensWithinClusterNoncriticalRegionClusterBalancedMeanPercent\%$
            & $\SimulatedWithoutLineOfSightIntrinsicLensWithinClusterNoncriticalRegionClusterBalancedMeanPercent\%$
            & $\SimulatedWithLineOfSightIntrinsicLensWithinClusterNoncriticalRegionClusterBalancedMeanPercent\%$
            & $\SimulatedWithCoreCollapsedSIDMIntrinsicLensWithinClusterNoncriticalRegionClusterBalancedMeanPercent\%$ \\
        \quad Boosted Lens
            & $\ObservedBoostedLensWithinClusterNoncriticalRegionClusterBalancedMeanPercent\%$
            & $\SimulatedWithoutLineOfSightBoostedLensWithinClusterNoncriticalRegionClusterBalancedMeanPercent\%$
            & $\SimulatedWithLineOfSightBoostedLensWithinClusterNoncriticalRegionClusterBalancedMeanPercent\%$
            & $\SimulatedWithCoreCollapsedSIDMBoostedLensWithinClusterNoncriticalRegionClusterBalancedMeanPercent\%$ \\
        \quad Inert Member
            & $\ObservedInertMemberWithinClusterNoncriticalRegionClusterBalancedMeanPercent\%$
            & $\SimulatedWithoutLineOfSightInertMemberWithinClusterNoncriticalRegionClusterBalancedMeanPercent\%$
            & $\SimulatedWithLineOfSightInertMemberWithinClusterNoncriticalRegionClusterBalancedMeanPercent\%$
            & $\SimulatedWithCoreCollapsedSIDMInertMemberWithinClusterNoncriticalRegionClusterBalancedMeanPercent\%$ \\
        \hline
    \end{tabular}
\end{table*}

\begin{figure*}[ht]
    \centering
    \includegraphics[width=\textwidth]{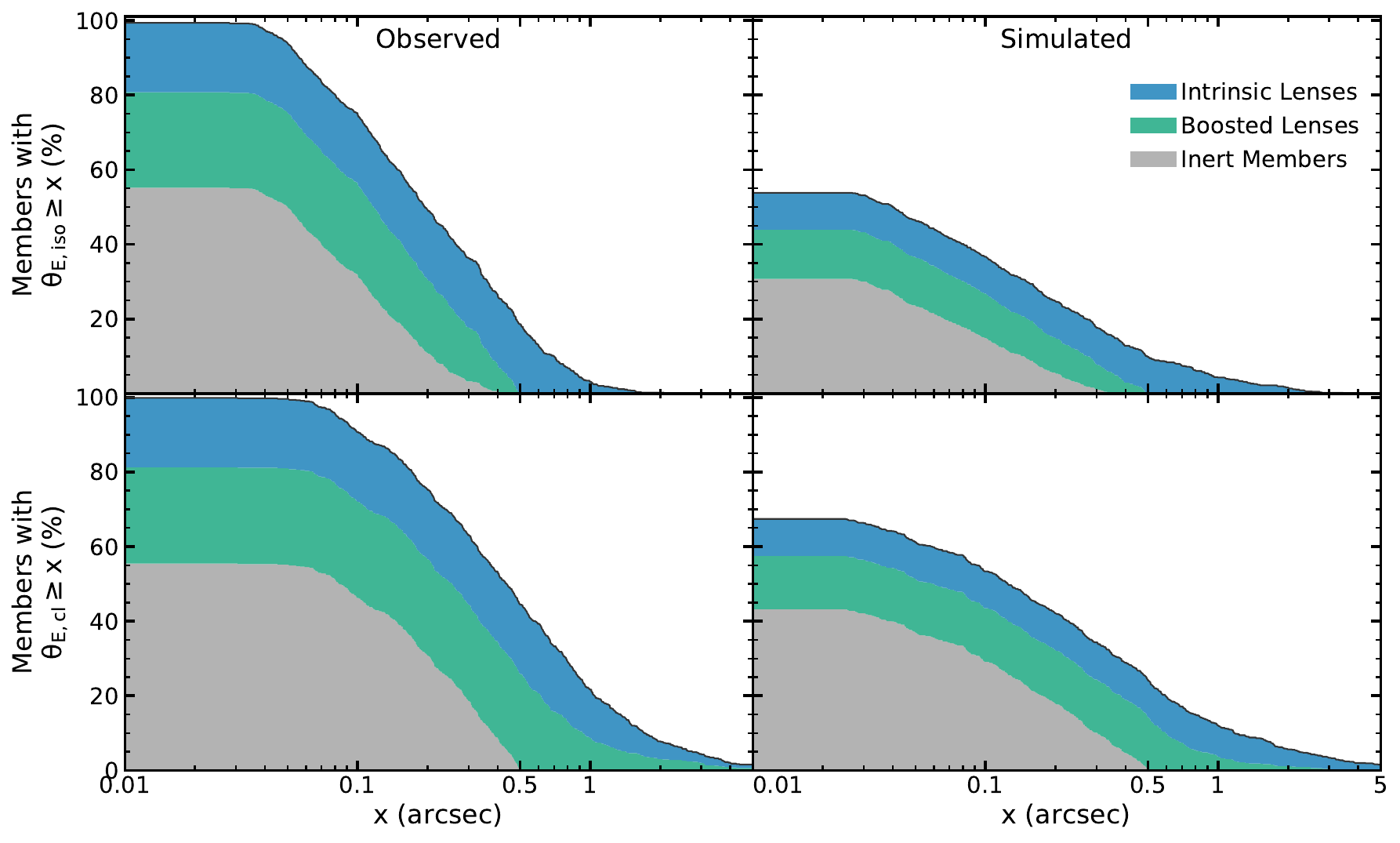}
\caption{\textbf{Cumulative member lensing strength by lensing class:} 
the left and right columns show the observed clusters and matched TNG-Cluster analogs, respectively. The top row shows the fraction of selected members with isolated Einstein radius 
$\theta_{E,\rm iso}\ge x$, normalized within that population. The bottom row shows the fraction with cluster-field Einstein radius $\theta_{E,\rm cl}\ge x$. Contributions are separated into Intrinsic Lenses, Boosted Lenses, and Inert Members. The reference scale $\theta_{\rm ref}=0.5 \arcsec$ marks the adopted GGSL threshold. Relative to the simulations, the observed clusters contain more members with large isolated lensing strength and a larger population that exceeds the GGSL threshold after the local cluster field is included. The class decomposition shows that this difference is driven primarily by the larger observed populations of Intrinsic and Boosted Lenses.}
    \label{fig:class_contributions}
\end{figure*}

Figure~\ref{fig:ratio_barcode} and Table~\ref{tab:type_fractions} show differences between observations and simulations in both the fraction of members in critical environments and the lensing strengths of those in subcritical environments. Among members in the cluster lensing-critical regions, the Major and Minor Perturber fractions in the simulations agree with the observations. The observed clusters, however, place a larger fraction of all members in these regions. Moreover, outside the cluster lensing-critical regions, the observed clusters contain larger fractions of Intrinsic and Boosted Lenses and fewer Inert Members. The cluster environment boosts more otherwise subcritical galaxies above the strong-lensing regime in the observed clusters than it does in the simulations. 
Figure~\ref{fig:class_contributions} extends this comparison across Einstein radius. The top row shows the member lensing strengths in isolation, while the bottom includes the local cluster field. At $x=0\farcs5$, the isolated fraction counts Intrinsic Lenses, and the cluster-field fraction counts members that meet the lensing criterion after including their environment.

The difference in lensing class is important because the overall abundance of substructure is similar in observations and simulations. The simulations do not primarily lack subhalos; instead, too many members remain lensing-inert even in environments where cluster convergence and shear should promote marginal galaxies above the strong-lensing threshold. GGSL therefore reveals a shortage of sufficiently dense members in the near-critical regime, rather than a simple deficit of substructure.

\subsection{Dependence on mass, radius, and line-of-sight structure}
\label{subsec:results_mass_radius_los}
\begin{figure*}[ht]
    \centering
    \includegraphics[width=\textwidth]{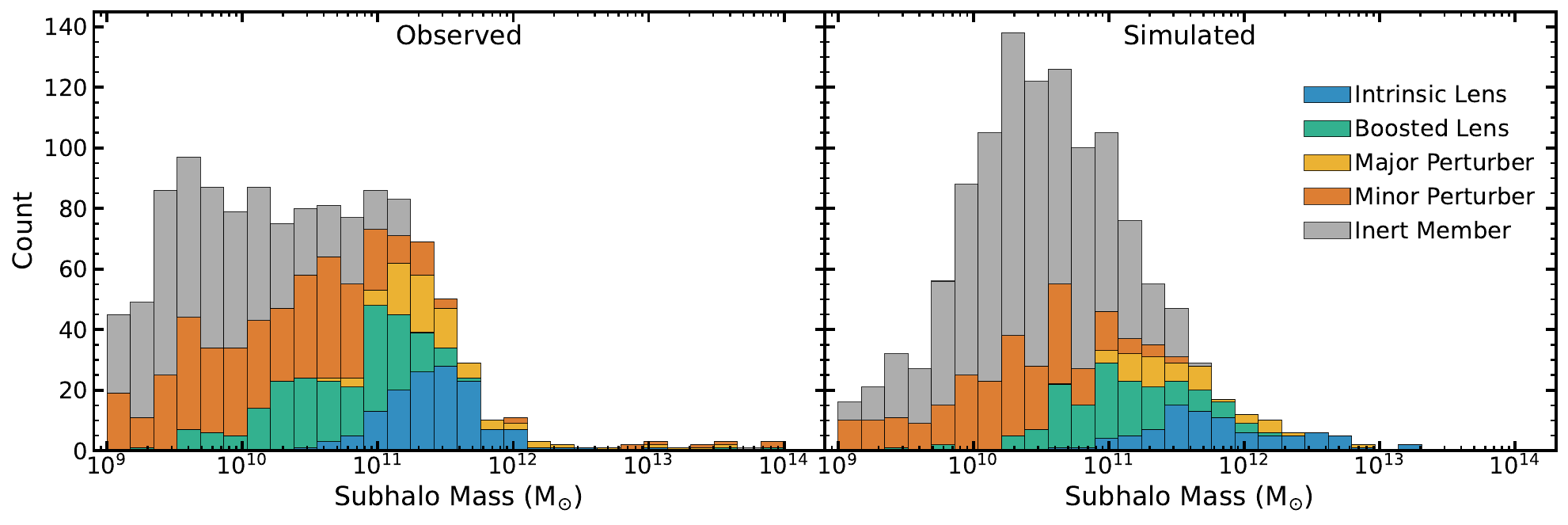}
\caption{\textbf{Subhalo mass distributions separated by lensing class:} the left panel shows the selected members of the six observed clusters, while the right panel shows one TNG-Cluster analog for each observed cluster, chosen as the closest match in primary Einstein radius $\theta_{\rm E,p}$. Histograms are divided by lensing class: Intrinsic Lenses, Boosted Lenses, Major Perturbers, Minor Perturbers, and Inert Members. The overall subhalo-mass distributions are similar in the observed and simulated samples, but their lensing classifications differ at fixed mass. Observed clusters contain a larger fraction of Intrinsic and Boosted Lenses, whereas simulated members of comparable mass are more often classified as Inert. This shows that the discrepancy is not primarily driven by a deficit of massive subhalos, but by differences in their lensing efficiency at fixed mass.}
    \label{fig:phase_mass_counts}
\end{figure*}

We next test whether these lensing-class differences persist at fixed subhalo mass and across cluster environment.

Figure~\ref{fig:phase_mass_counts} shows that the subhalo mass functions derived from the observationally constrained lens models and their simulated analogs are in very good agreement. However, the observed clusters have strong lenses across the entire mass range, whereas simulated members of comparable mass are more often inert. Dynamical mass alone is therefore not the deciding factor. At fixed mass, fewer simulated members retain enough projected mass in their inner few kiloparsecs to form a critical curve, either independently or with assistance from the cluster environment.

\begin{figure*}[ht]
    \centering
    \includegraphics[width=\textwidth]{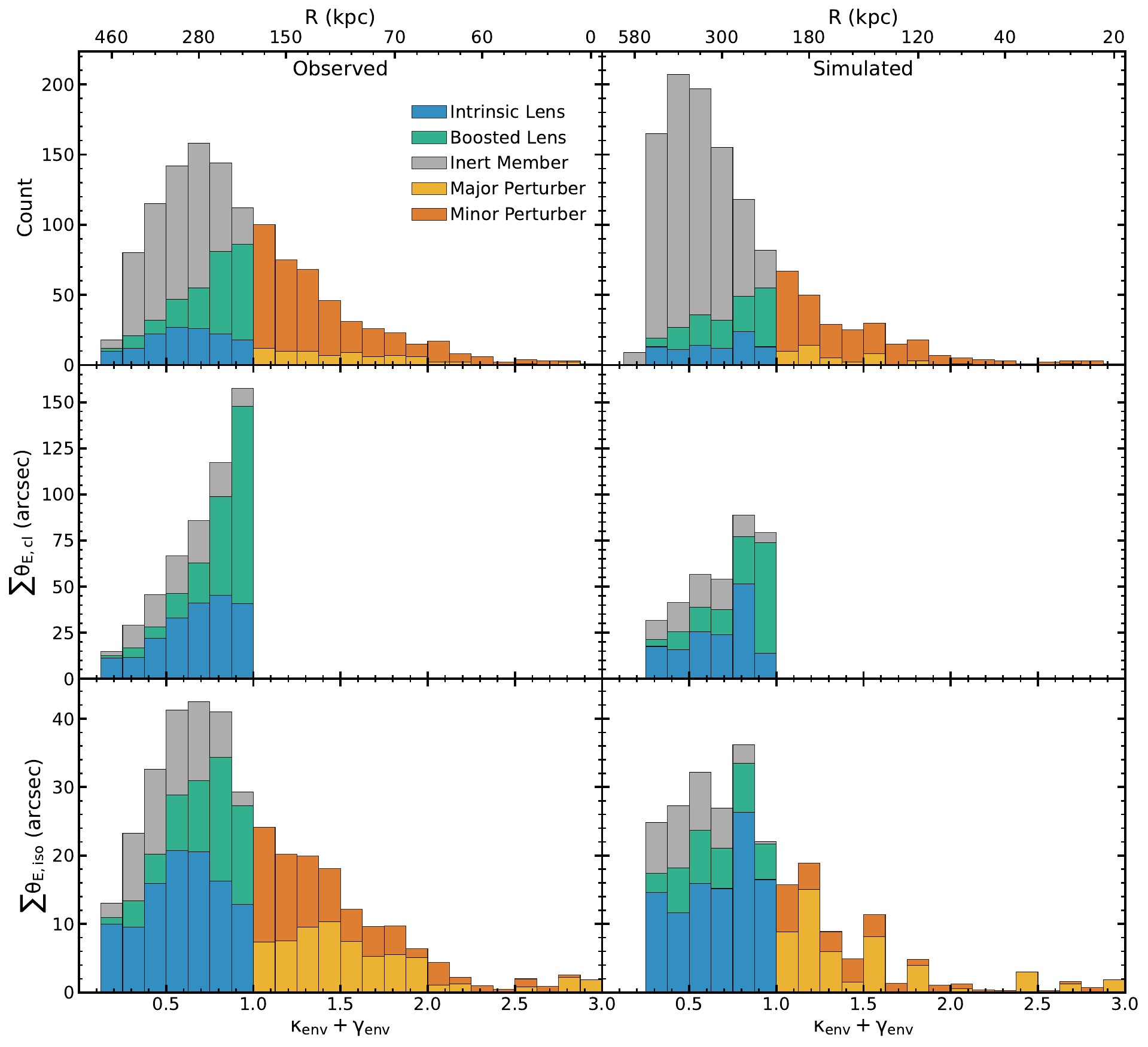}
\caption{\textbf{Member abundance and lensing strength as a function of local cluster environment:}
the left column shows the observed clusters and the right column their matched TNG-Cluster analogs. The top row gives the number of members in bins of environmental strength, parameterized by $\kappa_{\rm env}+\gamma_{\rm env}$, with colors indicating lensing class. The middle and bottom rows show, respectively, the summed cluster-field Einstein radii, $\theta_{E,\rm cl}$, and summed isolated Einstein radii, $\theta_{\rm E, iso}$, contributed by members in each bin. The upper horizontal axis gives the median projected cluster-centric radius of members in each environmental bin, illustrating the correspondence between stronger local lensing fields and smaller projected radii. Compared with the simulations, the observed clusters place more members in the strongest environments and contain greater galaxy-scale lensing strength across the subcritical regime, particularly from Intrinsic and Boosted Lenses.}
    \label{fig:phase_radial}
\end{figure*}

The distributions in Figure~\ref{fig:phase_radial} separate the contribution of the cluster environment from the lensing strength of the galaxies themselves. The observed clusters contain a larger fraction of members whose surrounding field is already critical. The combined Perturber fraction is \ObservedPerturbersClusterBalancedMeanPercent\% in the observations and \SimulatedWithoutLineOfSightPerturbersClusterBalancedMeanPercent\% in the simulated analogs, a difference of \ObservedMinusSimulatedWithoutLineOfSightPerturbersClusterBalancedPercentagePointDifference\ percentage points. This excess is present in five of the six observed clusters relative to the mean of their respective matched analog ensembles (see Table~\ref{tab:cluster_fractions} in the Appendix). However, the discrepancy persists where the environment is subcritical. There, the simulations contain fewer Intrinsic and Boosted Lenses and more Inert Members than their matched observed clusters (Figure~\ref{fig:ratio_barcode}). The overall lensing-class fractions, therefore, depend on both where member galaxies lie within the cluster field, which determines their environmental convergence and shear, and whether those galaxies are sufficiently dense to lens.

This environmental discrepancy has a clear spatial connection. The median projected radii on the top axes generally decrease toward stronger environmental fields, linking the shortage of simulated members in critical environments to the discrepancy in radial distributions reported by \citet{Natarajan+2026}. The middle and bottom rows of Figure~\ref{fig:phase_radial} show the discrepancy in lensing strength rather than member counts. Across subcritical environments associated with a range of projected radii, observed Intrinsic and Boosted Lenses contribute larger summed cluster-field Einstein radii than their simulated counterparts. Thus, the discrepancy persists even where environmental boosting is available.

\begin{figure}[ht]
    \centering
    \includegraphics[width=\linewidth]{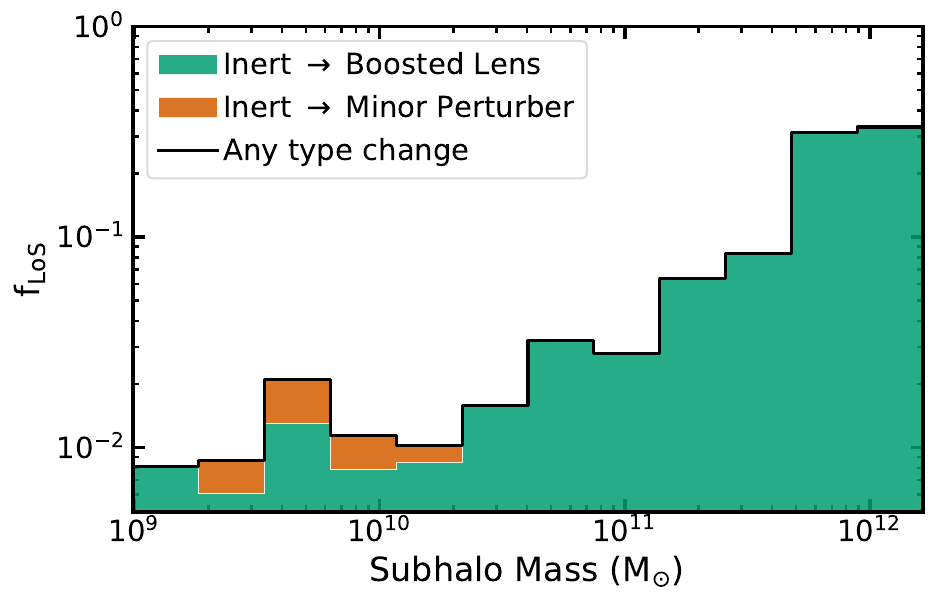}
    \caption{
    \textbf{Changes in lensing-class caused by line-of-sight structure:} colored regions show the cluster-balanced fraction of baseline Inert Members that become Boosted Lenses or Minor Perturbers after the line-of-sight planes are included. The black step curve gives the fraction undergoing any lensing-class change.
    }
    \label{fig:los_hist}
\end{figure}

Line-of-sight structure changes the classifications of $\SimulatedLineOfSightClassChangeClusterBalancedMeanPercent\%$ of selected simulated members. The unconditional Boosted Lens fraction increases from $\SimulatedWithoutLineOfSightBoostedLensClusterBalancedMeanPercent\%$ to $\SimulatedWithLineOfSightBoostedLensClusterBalancedMeanPercent\%$. Similarly, the change in the combined Perturber fraction is minimal, with an increase from $\SimulatedWithoutLineOfSightPerturbersClusterBalancedMeanPercent\%$ to $\SimulatedWithLineOfSightPerturbersClusterBalancedMeanPercent\%$. Both remain below the observed fractions of $\ObservedBoostedLensClusterBalancedMeanPercent\%$ and $\ObservedPerturbersClusterBalancedMeanPercent\%$, respectively. Line-of-sight structure can move members across nearby classification boundaries, as shown in Figure~\ref{fig:los_hist}, where for example up to $\sim 30\%$ of $10^{12}\,\Msun$ inert subhalos cross into boosted lenses. However, the over-dense line of sight neither supplies the missing galaxy-scale lensing strength nor reproduces the larger occupancy of cluster-critical regions seen in the observed sample.

\subsection{Phenomenological test motivated by core-collapsed SIDM}
\label{subsec:results_csidm}

\begin{figure*}[ht]
    \centering
    \includegraphics[width=\linewidth]
    {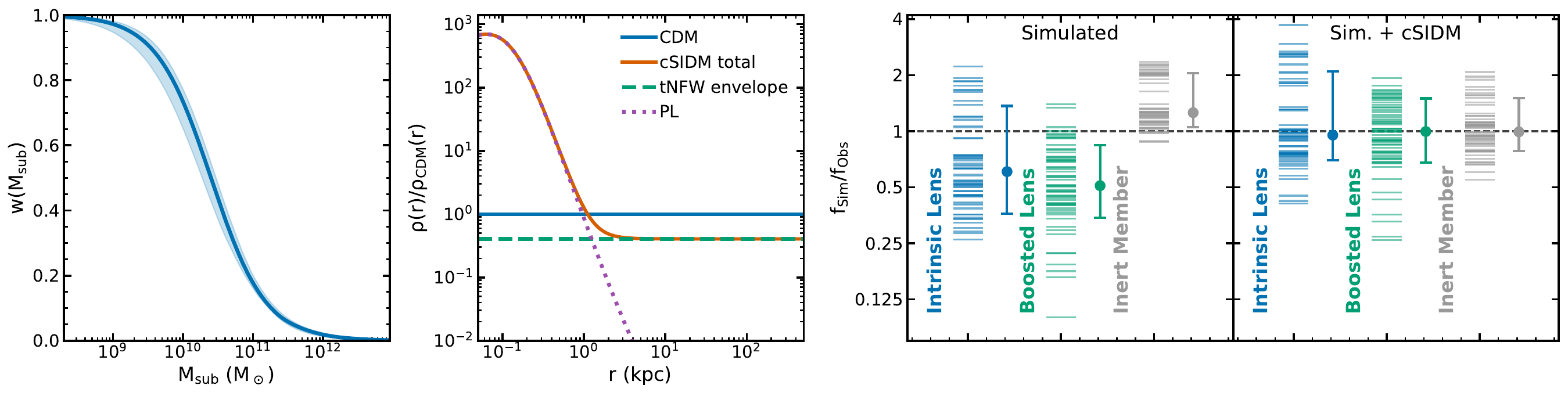}
\caption{The first panel shows the fraction of dark-matter mass transferred from the tNFW envelope to the Plummer component as a function of subhalo mass. The solid curve and shaded band show the median and 16th--84th percentiles from 1,000 bootstrap refits. The second panel shows the dark matter density profile of a representative member before and after the response. The final two panels compare the simulated and observed Intrinsic, Boosted, and Inert fractions, before and after the response. Each horizontal mark represents one analog relative to its matched observed cluster. Circles and capped intervals show the cluster-balanced median and 16th--84th percentiles; the dashed line marks agreement with the observations. Both calculations exclude line-of-sight structure. This phenomenological response mimics the central contraction expected from gravothermal core collapse but is not a calibrated SIDM prediction. We note that the low mass end of this plot ($\sim10^9$) is derived from the lowest subhalo mass that can be currently resolved in state-of-the-art lensing mass-models for observed clusters.}
    \label{fig:sidm}
\end{figure*}

The comparison above separates two contributions to the discrepancy. The observed clusters contain more members in the cluster lensing critical region, and their member galaxies are also stronger lenses outside the region. The latter difference points to greater projected mass within the inner few kiloparsecs, while leaving the total subhalo mass and tidal extent unchanged \citep{Chiang2026,Natarajan+2026}. Gravothermal core collapse in self-interacting dark matter can produce such compact inner profiles \citep{Balberg2002,Koda2011,Essig2019,Yang2023}. We therefore test whether a simple mass-dependent response that mimics core-collapsed SIDM can reproduce the lensing-type fractions outside the cluster-critical regions.

We apply the response only to satellite dark-matter components, modeled by a tNFW profile. For each satellite, we transfer a fraction of the finite dark-matter mass into a spherical Plummer component. We define $m_{\rm sub}\equiv\log(M_{\rm sub}/M_\odot)$ and $m_{\rm t}\equiv\log(M_{\rm t}/M_\odot)$, and write the transferred fraction as
\begin{equation}
w(M_{\rm sub}) = \left[
    1+\exp\left(\frac{m_{\rm sub}-m_{\rm t}}{\Delta_M}\right)
\right]^{-1},
\label{eq:csidm_response}
\end{equation}
where $M_{\rm t}$ is the transition mass and $\Delta_M$ sets its width in logarithmic mass. The transferred fraction approaches unity at low mass and zero at high mass, with $w=1/2$ at $M_{\rm sub}=M_{\rm t}$. Here, $M_{\rm sub}$ is the total bound Subfind mass used to set the response weight.

We transform each eligible dark-matter profile accordingly.
\begin{equation}
\rho_{\rm DM}'(r) = (1-w)\rho_{\rm tNFW}(r) +
    \frac{3wM_{\rm DM,tot}}{4\pi a^3} \left(1+\frac{r^2}{a^2}\right)^{-5/2},
\label{eq:csidm_profile}
\end{equation}
where $M_{\rm DM,tot}$ is the finite total mass of the original fitted tNFW profile, and $a$ is the Plummer scale shared by all transformed satellites. The residual tNFW envelope retains its original scale and truncation radii, while its density normalization decreases by a factor $1-w$. The Plummer component contains the remaining mass $wM_{\rm DM,tot}$, so the transformation conserves the total dark-matter mass. Stars, gas, black holes, member positions, and the smooth cluster component remain unchanged.

We fit $M_{\rm t}$, $\Delta_M$, and $a$ by minimizing the squared differences between the simulated and observed Intrinsic, Boosted, and Inert fractions across all clusters with equal weight assigned to each of the three conditional class fractions and to each of the six clusters. We estimate uncertainties from 1,000 bootstrap refits. We report the bootstrap medians and 16th--84th percentile intervals. We then recompute the critical curves, local environmental fields, lensing radii, and member classes, allowing members to change environmental regime. We find $M_{\rm t} =\CoreCollapsedSIDMResponseTransitionMass\,M_\odot$, $\Delta_M=\CoreCollapsedSIDMResponseLogMassWidth$ dex, and
$a=\CoreCollapsedSIDMPlummerScale$. 

\begin{figure*}[t]
\centering
\includegraphics[width=\textwidth]
{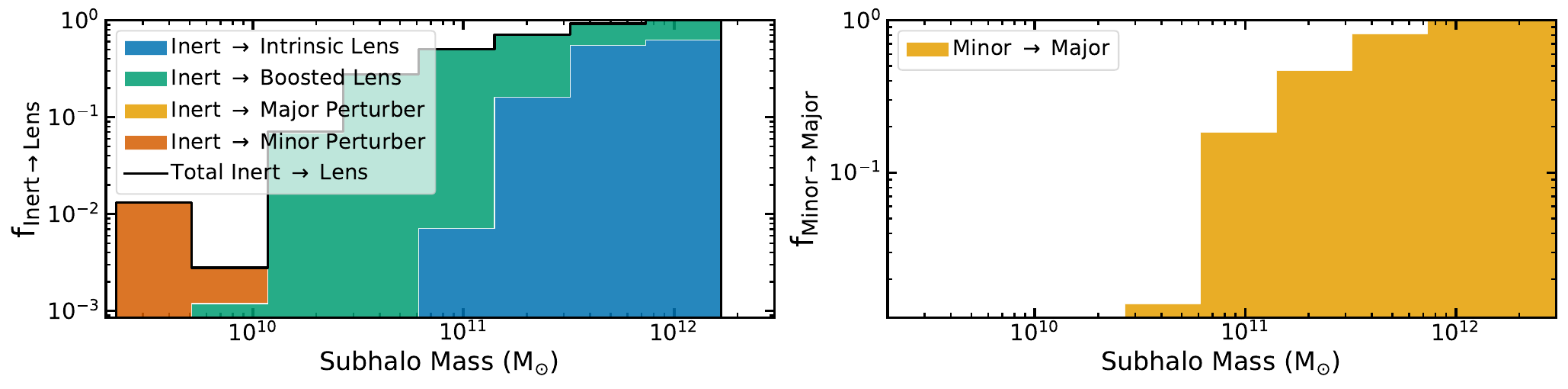}
\caption{
\textbf{Mass dependence of lensing class changes under the cSIDM response:} the left panel shows the fraction of baseline Inert Members reclassified as Intrinsic or Boosted Lenses after applying the cSIDM response. The colored regions separate the two final types, while the black line gives their combined fraction. The right panel shows the fraction of baseline Minor Perturbers promoted to Major Perturbers. Each fraction is normalized by its baseline population within the corresponding subhalo-mass bin. We exclude line-of-sight structure.}
\label{fig:csidm_activation_mass}
\end{figure*}

Figure~\ref{fig:csidm_activation_mass} shows how the resulting changes in lensing state depend on member mass. Figure~\ref{fig:sidm} shows that the response increases the dark-matter density on galaxy-lensing scales while preserving the total dark matter mass. It reduces the excess of Inert Members outside the cluster-critical regions and raises the Boosted Lens fraction toward the observations. Although member positions and the smooth cluster component remain fixed, the response changes the environmental fields contributed by neighboring members. The Perturber fraction can therefore change when we recompute the classifications. This experiment demonstrates that a mass-selective increase in central dark-matter density can reproduce the conditional lensing demographics without changing subhalo masses or truncation radii. Because these fractions were the fit target, the result is a sufficiency test rather than an independent constraint on SIDM. The response is phenomenological and should not be interpreted as a calibrated SIDM profile or a measurement of the scattering cross section.

\section{Discussion and Conclusions}
\label{sec:discussion}

GGSL combines two physically distinct ingredients: the intrinsic lensing strength of cluster member galaxies and the enhancement supplied by the surrounding cluster potential. GGSL occurs when an individual member produces a secondary critical curve, distinct from the cluster-scale critical curve. Its probability is determined by the total source-plane area enclosed by the corresponding secondary caustics. These galaxy-scale lenses generally lie where the smooth cluster potential provides substantial convergence and shear, thereby enhancing their lensing strength, but is not itself critical. A measurement of the total GGSL probability alone, therefore, cannot determine whether a simulated cluster fails because its members are too diffuse, because too few members occupy favorable locations, or because both effects occur simultaneously.

Our member-by-member decomposition presented in this work separates these contributions by measuring the lensing strength of every galaxy in isolation and after embedding it in its local cluster convergence and shear field. The classification is deliberately hierarchical. We first determine whether a member lies where the surrounding cluster field is already critical, perturbing the existing cluster-scale critical curve rather than generating a distinct secondary critical curve. Such members are classified as \emph{Major Perturbers} when they meet the adopted $\theta_{\rm ref}=0.5^{\prime\prime}$ threshold in isolation and as \emph{Minor Perturbers} when they do not. Members outside the cluster-critical regions are instead classified according to the comparison between their isolated and environmentally enhanced Einstein radii. \emph{Intrinsic Lenses} exceed the threshold without assistance from the host cluster, \emph{Boosted Lenses} are subcritical in isolation but cross the threshold after the local cluster field is included, and \emph{Inert Members} remain below the threshold even with environmental assistance. These five lensing classes are therefore not simply bins in subhalo mass. They encode both the position of each member and the projected density of its inner mass distribution.

This separation reveals a clear spatial discrepancy. The observed clusters place $\ObservedClusterCriticalRegionClusterBalancedMeanPercent\%$ of their selected members in regions where the cluster potential is already critical, whereas the corresponding fraction in the TNG-Cluster analogs is only $\SimulatedWithoutLineOfSightClusterCriticalRegionClusterBalancedMeanPercent\%$. This offset is not driven by a single system: in five of the six observed clusters, the critical-region occupancy exceeds the median of the corresponding matched TNG-Cluster analog ensemble. Conditional on lying in these regions, however, the composition of the population is similar: Major Perturbers constitute $\ObservedMajorPerturberWithinClusterCriticalRegionClusterBalancedMeanPercent\%$ of the observed Perturbers and $\SimulatedWithoutLineOfSightMajorPerturberWithinClusterCriticalRegionClusterBalancedMeanPercent\%$ of the simulated Perturbers, with Minor Perturbers dominating both samples. Thus, the overall shortage of simulated Perturbers does not arise primarily because galaxies already situated on the cluster critical structure have drastically incorrect isolated lensing strengths. It arises because too few simulated members occupy the region jointly defined by the member distribution and the macro-critical curves. This finding connects the GGSL discrepancy directly to the observed--simulated difference in the radial distribution of cluster substructure. Altering the internal profiles of individual members cannot, by itself, correct this occupancy difference because it neither moves the galaxies nor substantially relocates the smooth cluster critical structure.

A second and independent discrepancy appears outside the cluster-critical regions, where distinct galaxy-scale secondary critical curves can form. The observed clusters contain $\ObservedIntrinsicLensWithinClusterNoncriticalRegionClusterBalancedMeanPercent\%$ Intrinsic Lenses, $\ObservedBoostedLensWithinClusterNoncriticalRegionClusterBalancedMeanPercent\%$ Boosted Lenses, and $\ObservedInertMemberWithinClusterNoncriticalRegionClusterBalancedMeanPercent\%$ Inert Members. The corresponding TNG-Cluster fractions are $\SimulatedWithoutLineOfSightIntrinsicLensWithinClusterNoncriticalRegionClusterBalancedMeanPercent\%$, $\SimulatedWithoutLineOfSightBoostedLensWithinClusterNoncriticalRegionClusterBalancedMeanPercent\%$, and $\SimulatedWithoutLineOfSightInertMemberWithinClusterNoncriticalRegionClusterBalancedMeanPercent\%$, respectively. Figure~\ref{fig:class_contributions} shows how the observed and simulated member fractions change with the Einstein radius. The most pronounced difference is, therefore, not simply a shortage of intrinsically critical galaxies but the absence of a substantial intermediate population lying sufficiently close to criticality to be promoted into the strong-lensing regime by the host cluster. In the observations, the subcritical convergence and shear of the cluster environment repeatedly convert marginal galaxies into Boosted Lenses. In the simulations, many members remain inert even where environmental boosting should be most effective.

This internal structure discrepancy cannot be attributed to a simple deficit of massive subhalos. The observed and simulated subhalo mass functions are broadly consistent; yet, simulated members of comparable total mass are much more likely to remain lensing-inert. The relevant difference must therefore reside in how mass is distributed within the inner few kiloparsecs rather than in the total bound mass alone. Nor is the result confined to the innermost cluster core. The observed sample sustains a population of lensing-relevant members over a broad range of projected radius, whereas the simulated analogs remain increasingly dominated by Inert Members. Line-of-sight structure produces occasional transitions across nearby class boundaries but changes the aggregate fractions only modestly. It therefore neither supplies the missing central surface density nor accounts for the low occupancy of the simulated cluster-critical regions.

The combined inner and outer lensing constraints sharply restrict viable explanations. GGSL requires a greater projected density on galaxy-lensing scales, while independent measurements indicate that the truncation radii and outer tidal extents of observed cluster members remain broadly compatible with collisionless-CDM predictions. The solution cannot therefore be a uniform increase in subhalo density or mass at all radii. Instead, it must preferentially rearrange existing mass toward the center while retaining an approximately CDM-like outer extent. Numerical disruption, force softening, baryonic feedback, or deficiencies in the modeled stellar distributions could contribute to the low central densities of simulated members. Alternatively, late-stage gravothermal evolution in self-interacting dark matter provides a physical mechanism capable of producing compact central cusps without necessarily requiring excessive mass loss at the tidal boundary, particularly when the interaction cross section decreases with velocity.

Our phenomenological contraction experiment tests the sufficiency of such an inward redistribution. Transferring part of the dark-matter mass into a compact Plummer component increases the lensing strength of simulated members while conserving their finite dark-matter mass. The resulting fractions outside the cluster-critical regions approach those of the observed sample, and some Minor Perturbers become Major Perturbers as their isolated lensing strength increases. The experiment therefore cleanly separates the two effects: central contraction can alleviate the structural discrepancy, but it cannot repair the spatial-occupancy discrepancy.

This exercise should not be interpreted as a calibrated SIDM prediction or as a measurement of the dark-matter scattering cross section. The contraction was fitted to the same conditional lensing fractions used to assess its success and is consequently a sufficiency test rather than an independent inference. It does not follow the time-dependent evolution of individual satellites, account for the diversity of their orbital and stripping histories, or predict which members should have entered the core-collapse phase. Nevertheless, it establishes an important physical result: redistributing existing dark matter inward can generate the missing population of environmentally boosted secondary lenses without increasing total subhalo masses or violating the observed outer tidal extents.

Several limitations of the present comparison should be kept in view. The analysis applies to the strong-lensing region rather than to the complete member population out to $R_{200}$. The observational inference inherits the assumptions of parametric cluster lens models and luminosity-based member scaling relations \citep{Meneghetti2017:HFFComparison, Granata2023}, while the simulated calculation relies on analytic representations fitted to resolved dark-matter, stellar, and gaseous components. The selected clusters are necessarily systems with high-quality strong-lensing constraints and do not constitute an unbiased sample of the full cluster population. Cluster orientation, dynamical state, merger geometry, and the covariance between the smooth potential and member-galaxy parameters could alter the detailed demographics. These effects may broaden the cluster-to-cluster distribution, but the persistence of both the occupancy and structural differences across the matched ensembles suggests that neither can be reduced to a single modeling choice.

To summarize, our key findings are:

\textbf{(i) Member lensing class.}
Comparing each member's Einstein radius in isolation and in the local cluster field separates Intrinsic Lenses, Boosted Lenses, and Inert Members. Members that lie where the large-scale cluster potential is already critical are classified as Major or Minor Perturbers, according to whether they can lens in isolation.

\textbf{(ii) Occupancy of the cluster-critical regions.}
The observed clusters contain more members in cluster-critical regions than the TNG-Cluster analogs. Because the Major-to-Minor Perturber ratio within those regions agrees, this difference primarily reflects the relative placement of cluster members with respect to the macro-critical structure.

\textbf{(iii) Member structure outside the cluster-critical regions.}
Too many simulated members remain Inert where the cluster field could otherwise promote them into the strong-lensing regime. Their inner projected densities, rather than their total masses alone, limit their contribution to GGSL. Inclusion of line-of-sight structure has minimal effect on their lensing strength.

\textbf{(iv) Effect of central contraction.}
A mass-conserving increase in central dark-matter density reproduces the conditional fractions of Intrinsic, Boosted, and Inert Members while preserving the truncation radius. However, it does not reproduce the observed occupancy of cluster lensing-critical regions. Central contraction can therefore account for the difference in inner structure and resulting lensing of the members, but the spatial discrepancy requires a separate explanation.

These two requirements connect our TNG-Cluster comparison to the broader difficulty that cosmological hydrodynamical simulations have in reproducing the observed GGSL signal \citep{Meneghetti2020,Meneghetti2022,Meneghetti2023}. Matching the number of cluster galaxies or subhalo masses is not sufficient. Simulations must also place enough members where the cluster potential is already critical or near critical and also preserve the compact inner densities that allow those members to remain efficient lenses after infall and tidal processing.


Future work should replace the phenomenological contraction with forward predictions from high-resolution simulations that resolve the coupled evolution of stars and dark matter in cluster satellites. Matched CDM and velocity-dependent SIDM zoom simulations would allow the five lensing classes to be followed as functions of infall time, pericentric distance, tidal mass loss, stellar compactness, and gravothermal phase. Applying the same lensing pipeline directly to mock observations would determine whether the required central enhancement can be produced while simultaneously retaining the observed truncation radii and radial distribution. On the observational side, enlarging the cluster sample with uniform HST and JWST lens models, member stellar kinematics, and spectroscopic redshifts will permit the lensing-class fractions to be measured as functions of cluster mass, dynamical state, and projected radius. The full distribution of isolated lensing strength and environmental criticality deficit, rather than only the five discrete lensing classes, may ultimately provide a sensitive empirical map of subhalo structural evolution within clusters.

Taken together, our results show that the GGSL discrepancy is not a one-parameter failure that can be explained solely by changing subhalo abundance or concentration. The observed clusters contain more galaxies in the locations where the macro potential is critical, while outside those regions, their members retain greater central lensing strength and generate a larger total area of secondary critical curves. Any successful model must therefore reproduce both the spatial distribution of cluster members and the internal structure that allows marginal galaxies to respond to the subcritical magnification supplied by the cluster environment. This two-component interpretation provides a more precise target for simulations and clarifies why matching the subhalo mass function alone is insufficient to reproduce the observed strong-lensing phenomenology.

\vspace{-0.1cm} 
\begin{acknowledgments}
I.D. acknowledges support from NASA under award No. 80NSSC25K0311 under the NASA FINESST program.
P.N. gratefully acknowledges funding from the Department of Energy grant DE-SC0017660; support from the John Templeton Foundation via grant 126613 and from the NASA GLIMPSE JWST GO-03293.026 for this work.
\end{acknowledgments}


\appendix

\section{Comparison with radii measured from critical curves}
\label{appendix:hessian}

We validate the Einstein radii defined in Section~\ref{subsec:einstein_radii} with radii measured from the areas enclosed by tangential critical curves. This comparison tests the approximation used to include the surrounding field while keeping the member profiles and environmental convergence and shear fixed.

For each member, we calculate the Hessian of its lensing potential,
\begin{equation}
H_{ij,k}(\boldsymbol{\theta})
=\frac{\partial^2\psi_k}{\partial\theta_i\partial\theta_j},
\end{equation}
where $i,j\in\{x,y\}$. We represent the surrounding field, excluding the member itself, by the constant environmental Hessian
\begin{equation}
\overline{\mathbf{H}}_{{\rm env},k}=
\begin{pmatrix}
\kappa_{\rm env}+\gamma_{1,\rm env} & \gamma_{2,\rm env}\\
\gamma_{2,\rm env} & \kappa_{\rm env}-\gamma_{1,\rm env}
\end{pmatrix},
\end{equation}
using the convergence and shear averaged around the member in the main analysis. The lensing Jacobian is
\begin{equation}
\mathbf{A}(\boldsymbol{\theta})
=\mathbf{I}-\mathbf{H}_k(\boldsymbol{\theta})-\overline{\mathbf{H}}_{{\rm env},k},
\end{equation}
where $\mathbf{I}$ is the identity matrix.
We set $\overline{\mathbf{H}}_{{\rm env},k}=0$ when measuring $\theta_{\rm E,iso}$ and retain it when measuring $\theta_{\rm E,cl}$. In each case, we locate the tangential critical curve where
\begin{equation}
\lambda_{\rm t}=1-\kappa-\sqrt{\gamma_1^2+\gamma_2^2}=0,
\end{equation}
with $\kappa$ and the shear components evaluated from the combined Hessian for the corresponding lens configuration. Adding the shear components before taking their magnitude retains the relative orientation of the member and environmental shear. This calculation does not assume an isothermal member profile.

We find curves with one crossing along every sampled direction and a connected interior. For these curves, the enclosed area and effective Einstein radius are
\begin{align}
A_{\rm crit}
&=\frac{1}{2}\int_0^{2\pi}r_{\rm crit}^{\,2}(\phi)\,{\rm d}\phi,\\
\theta_{\rm E}^{\rm H}
&=\sqrt{\frac{A_{\rm crit}}{\pi}},
\end{align}
where $r_{\rm crit}(\phi)$ is the distance from the member center
to the curve.

We assign zero when no critical curve is detected within the sampled range. As in the main analysis, we do not assign a cluster-field radius where the environmental field is already critical.

\begin{figure*}[t]
    \centering
    \includegraphics[width=\linewidth]
        {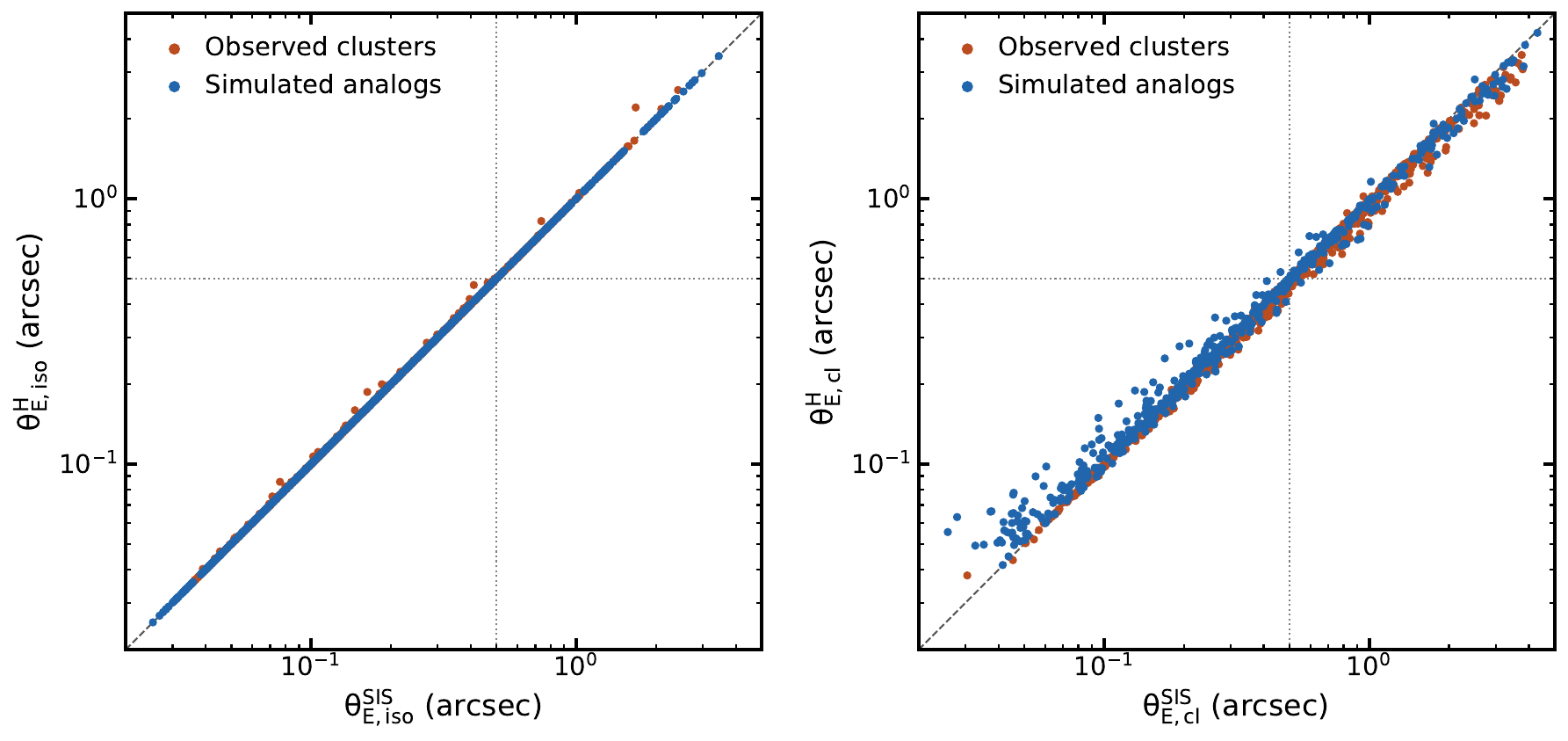}
\caption{\textbf{Validation of the Einstein radii using the SIS approximation:} the horizontal axes show the effective Einstein radii from the enclosed-convergence method adopted in the main text. The vertical axes show radii measured from the areas enclosed by critical curves located using the Hessian. The left panel shows members in isolation, and the right panel includes the environmental field. Orange points show observed members, and blue points show members of the simulated analogs without added line-of-sight structure. Dashed lines indicate equality. Dotted lines mark the reference Einstein radius $\theta_{\rm E,ref}=0\farcs5$ used to distinguish Inert Members, Boosted Lenses, and Intrinsic Lenses.}
    \label{fig:einstein_radius_method_comparison}
\end{figure*}

Figure~\ref{fig:einstein_radius_method_comparison} shows the six observed clusters and their six selected CDM analogs. The isolated radii agree closely, while including the surrounding field produces larger differences. The environmental correction is exact for a singular isothermal sphere in a uniform field, whereas the Hessian calculation uses the member's profile.

We quantify these differences using $100|\theta_{\rm E}/\theta_{\rm E,H}-1|$. Across the observed, CDM, cSIDM, and CDM plus line-of-sight populations, the median difference in cluster-field radii ranges from \HessianRadiusComparisonMedianDifferenceMin\% to \HessianRadiusComparisonMedianDifferenceMax\%.

Finally, we substitute the Hessian-based radii into the classification while keeping the environmental criterion fixed. The class assignments agree
in \HessianRadiusComparisonClassificationAgreementMin--%
\HessianRadiusComparisonClassificationAgreementMax\% of cases. All changes are between Boosted Lenses and Inert Members. These results support retaining the enclosed-convergence definition for the classification.

\section{Dependence on the reference radius}
\label{appendix:reference_radius}

We test whether the differences in lensing-class fractions are driven by the adopted reference radius of $0\farcs5$ by repeating the classification with $\theta_{\rm ref}=1''$. We retain the same observed members and simulated analogs, changing only the Einstein radius that separates Inert members from Lenses. The environmental classification remains unchanged.

Increasing the reference radius reduces the number of Intrinsic and Boosted Lenses that satisfy the lensing criteria, as expected from Figure~\ref{fig:class_contributions}. The smaller counts produce broader distributions of simulated-to-observed ratios for Intrinsic Lenses and Major Perturbers. Despite this increased scatter, the main trends persist - the simulations retain a lower Boosted Lens fraction and a higher Inert Member fraction. These differences are therefore not specific to the adoption of the Einstein radius reference of $0\farcs5$.

\begin{figure*}[ht]
    \centering
    \includegraphics[width=\textwidth]{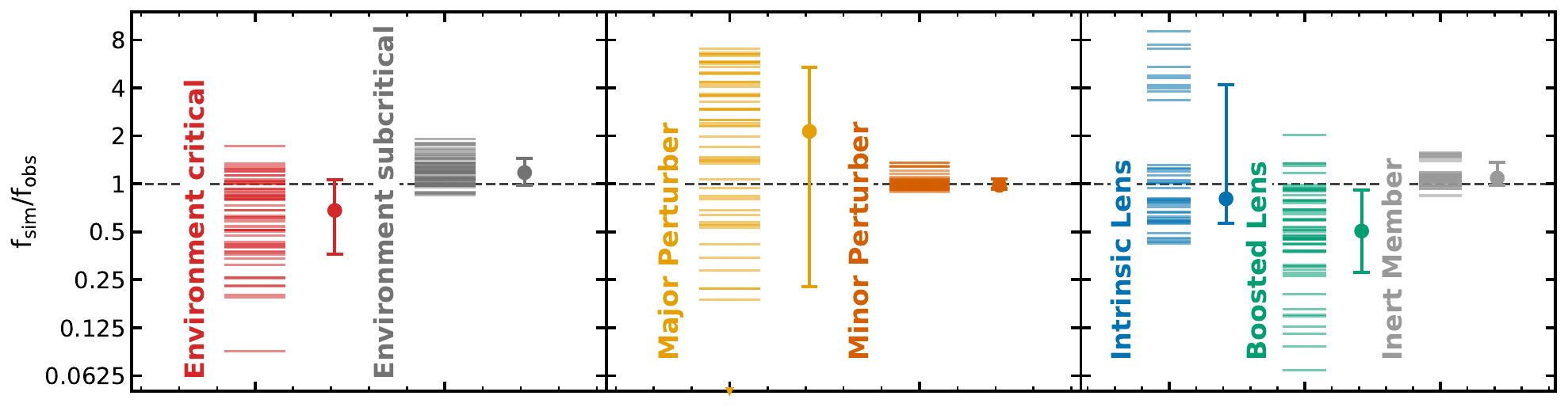}
    \caption{\textbf{Lensing-class fractions with a reference radius of $1''$.}
    As in Figure~\ref{fig:ratio_barcode}, marks show matched simulated-to-observed fraction ratios, and circles and intervals show the median and 16th--84th percentiles. Downward triangles indicate zero ratios. Increasing the reference radius broadens the ratio distributions for Intrinsic Lenses and Major Perturbers, while the simulated deficit of Boosted Lenses and excess of Inert Members persist.}
    \label{fig:ratio_barcode_theta_ref_1p0}
\end{figure*}

\section{Lensing-class fractions by cluster}
\label{appendix:cluster_fractions}

Table~\ref{tab:cluster_fractions} shows the environmental occupancy and conditional lensing-class fractions separately for each observed cluster and its matched simulated analogs.

\begin{table*}[ht!]
\centering
\caption{\textbf{Environmental occupancy and conditional lensing-class fractions by cluster.}
All entries are percentages. The first two columns of fractions use all selected members as the denominator. Major and Minor Perturber fractions are normalized within the critical environment; Intrinsic Lens, Boosted
Lens, and Inert Member fractions are normalized within the subcritical environment. Entries denoting simulations give the mean of the ten analog fractions per observed cluster, with equal weight assigned to each analog.}
\label{tab:cluster_fractions}
\begin{tabular}{llrrrrrrr}
\hline\hline
 & & \multicolumn{2}{c}{All members} & \multicolumn{2}{c}{Critical environment} & \multicolumn{3}{c}{Subcritical environment} \\
\cline{3-4}\cline{5-6}\cline{7-9}
Cluster & Sample & Critical & Subcritical & Major & Minor & Intrinsic & Boosted & Inert \\
\hline
Abell 2744 & Observed & 13.3 & 86.7 & 53.3 & 46.7 & 36.7 & 30.6 & 32.7 \\
 & Simulated & 15.2 & 84.8 & 23.3 & 76.7 & 16.9 & 16.1 & 67.0 \\
 & Sim. + LoS & 16.0 & 84.0 & 22.8 & 77.2 & 17.0 & 17.2 & 65.8 \\
 & Sim. + cSIDM & 15.3 & 84.7 & 30.1 & 69.9 & 29.6 & 31.5 & 38.9 \\
\hline
Abell 370 & Observed & 50.2 & 49.8 & 22.2 & 77.8 & 20.9 & 40.3 & 38.8 \\
 & Simulated & 18.6 & 81.4 & 14.7 & 85.3 & 8.5 & 10.7 & 80.8 \\
 & Sim. + LoS & 19.6 & 80.4 & 15.7 & 84.3 & 8.3 & 11.5 & 80.1 \\
 & Sim. + cSIDM & 18.8 & 81.2 & 20.6 & 79.4 & 14.4 & 19.7 & 65.9 \\
\hline
Abell S1063 & Observed & 37.6 & 62.4 & 20.3 & 79.7 & 18.7 & 17.1 & 64.2 \\
 & Simulated & 24.9 & 75.1 & 18.4 & 81.6 & 10.7 & 14.8 & 74.6 \\
 & Sim. + LoS & 25.4 & 74.6 & 18.5 & 81.5 & 10.6 & 16.2 & 73.2 \\
 & Sim. + cSIDM & 24.9 & 75.1 & 23.2 & 76.8 & 18.5 & 22.8 & 58.7 \\
\hline
MACS J0416 & Observed & 35.6 & 64.4 & 12.5 & 87.5 & 10.0 & 20.8 & 69.2 \\
 & Simulated & 22.4 & 77.6 & 16.0 & 84.0 & 11.3 & 12.7 & 75.9 \\
 & Sim. + LoS & 23.6 & 76.4 & 16.3 & 83.7 & 11.0 & 13.5 & 75.4 \\
 & Sim. + cSIDM & 22.4 & 77.6 & 20.6 & 79.4 & 18.0 & 22.9 & 59.2 \\
\hline
MACS J1206 & Observed & 31.5 & 68.5 & 11.4 & 88.6 & 17.8 & 20.4 & 61.8 \\
 & Simulated & 22.8 & 77.2 & 15.4 & 84.6 & 9.7 & 12.2 & 78.0 \\
 & Sim. + LoS & 24.0 & 76.0 & 16.2 & 83.8 & 9.4 & 13.5 & 77.1 \\
 & Sim. + cSIDM & 23.0 & 77.0 & 18.8 & 81.2 & 14.8 & 23.3 & 61.9 \\
\hline
PSZ1 G311 & Observed & 34.0 & 66.0 & 8.8 & 91.2 & 7.6 & 25.8 & 66.7 \\
 & Simulated & 25.5 & 74.5 & 17.2 & 82.8 & 11.8 & 14.7 & 73.5 \\
 & Sim. + LoS & 26.9 & 73.1 & 18.0 & 82.0 & 11.5 & 16.2 & 72.3 \\
 & Sim. + cSIDM & 25.6 & 74.3 & 20.7 & 79.3 & 18.2 & 26.3 & 55.5 \\
\hline
Mean across clusters & Observed & 33.7 & 66.3 & 21.4 & 78.6 & 18.6 & 25.8 & 55.6 \\
 & Simulated & 21.6 & 78.4 & 17.5 & 82.5 & 11.5 & 13.5 & 75.0 \\
 & Sim. + LoS & 22.6 & 77.4 & 17.9 & 82.1 & 11.3 & 14.7 & 74.0 \\
 & Sim. + cSIDM & 21.7 & 78.3 & 22.4 & 77.6 & 18.9 & 24.4 & 56.7 \\
\hline
\end{tabular}

\small
\setlength{\tabcolsep}{4pt}
\renewcommand{\arraystretch}{1.12}

\par\smallskip
\begin{minipage}{\textwidth}
\footnotesize
\textit{Notes.}
Critical denotes a locally tangentially critical or supercritical environment with the target member removed, $\lambda_{t,\rm env}\leq0$; subcritical denotes $\lambda_{t,\rm env}>0$.
The baseline Simulation (TNG-Cluster) sample excludes line-of-sight structure.
Simulation + LoS includes the adopted line-of-sight planes,  while Simulation + cSIDM applies the mass-conserving central contraction without line-of-sight structure. The final block weights the six clusters equally.
\end{minipage}
\end{table*}

\clearpage
\bibliographystyle{aasjournalv7}
\bibliography{refs}{}

\end{document}